\documentclass[a4paper,10pt]{article}
\pdfoutput=1
\usepackage{jcappub}
\usepackage{mathtools}
\usepackage{ulem}
\usepackage{bm}
\usepackage{xcolor}

\usepackage [latin1]{inputenc}

\renewcommand{\thesection}{\arabic{section}}

\def\laq{~\raise 0.4ex\hbox{$<$}\kern -0.8em\lower 0.62ex\hbox{$\sim$}~}
\def\gaq{~\raise 0.4ex\hbox{$>$}\kern -0.7em\lower 0.62ex\hbox{$\sim$}~}

\def\beq{\begin{equation}}
\def\eeq{\end{equation}}
\def\bea{\begin{eqnarray}}
\def\eea{\end{eqnarray}}

\def \fp {{\dot{\phi}}}

\def \pa {\partial}
\def \ra {\rightarrow}
\def \ti {\widetilde}

\def \ls {\lambda_{\rm s}}

\def \da {\delta}
\def \b {\beta}
\def \a {\alpha}
\def \ap {\alpha^{\prime}}

\def \ga {\gamma}

\def \da {\delta}
\def \ep {\epsilon}

\def \om {\omega}

\def \fpp {\ddot{\phi}}

\def \fb {\overline \phi}
\def \fbp {\dot{\fb}}
\def \fbpp {\ddot{\fb}}

\title{From the string  vacuum to FLRW or de Sitter via $\alpha'$ corrections}

\author[a]{P. Conzinu,}
\author[b]{G. Fanizza,}
\author[c]{M. Gasperini,}
\author[c]{E. Pavone,}
\author[c]{L. Tedesco}
\author[d]{and G. Veneziano}

\affiliation[a]{Dipartimento di Fisica, Universit\`a di Pisa, Largo B. Pontecorvo 3, 56127 Pisa, 
Italy,\\
and Istituto Nazionale di Fisica Nucleare, Sezione di Pisa, Italy}
\affiliation[b]{Instituto de Astrofis\'ica e Ci\^encias do Espa\c{c}o,
Faculdade de Ci\^encias da Universidade de Lisboa,
Edificio C8, Campo Grande, {P-1749-016} Lisbon, Portugal}
\affiliation[c]{
Dipartimento di Fisica, Universit\`a di Bari, 
Via G. Amendola 173, 70126 Bari, Italy,\\
and Istituto Nazionale di Fisica Nucleare, Sezione di Bari, Italy
}
\affiliation[d]{CERN, Theory  Department, CH-1211 Geneva 23, Switzerland,\\
and Coll\`ege de France, 11 Place M. Berthelot, 75005 Paris, France\\}

\emailAdd{pietro.conzinu@phd.unipi.it}
\emailAdd{gfanizza@fc.ul.pt}
\emailAdd{gasperini@ba.infn.it}
\emailAdd{eliseo.pavone@ba.infn.it}
\emailAdd{luigi.tedesco@ba.infn.it}
\emailAdd{gabriele.veneziano@cern.ch}

\abstract{We first make more precise a recent ``Hamiltonian" reformulation of the Hohm-Zwiebach approach to the tree-level, $O(d,d)$-invariant string cosmology equations at all orders in the $\ap$ expansion, and recall how it allows to give a simple characterization of  a large class of  cosmological scenarios connecting, through a non-singular bounce, two duality-related perturbative solutions at early and late times. We then discuss the effects of adding to the action a non-perturbative, $O(d,d)$-breaking, dilaton potential $V(\phi)$. The resulting cosmological solutions, assumed to approach at early times the perturbative string vacuum (with vanishing curvature and string coupling),   can stabilize the dilaton at late times and simultaneously approach   either a matter-dominated FLRW cosmology or a de-Sitter-like inflationary phase, depending on initial conditions and on the properties of $V(\phi)$ at moderate-coupling. We also identify a general mechanism for generating isotropic late-time attractors from a large basin of anisotropic initial conditions.}

\keywords{string theory and cosmology, cosmology of theories beyond the SM, initial conditions and eternal universe
 
\vskip18pt 

\noindent{\bfseries\large\sffamily{Preprints:}} BA-TH/808-23, CERN-TH-2023-163
}

\begin{document}
\maketitle
\vskip 0.8 cm

\section{Introduction} 
\label{sec1}

It is well-known \cite{Meissner:1991zj,Meissner:1991ge,1b,Gasperini:1993hu} that the tree-level equations of string cosmology in $d$ spatial dimensions, provided they possess $d$ abelian isometries\footnote{The generalization to $d'$ abelian isometries with $ d' < d$ has been discussed in \cite{Maharana:1992my}.}, are invariant under a continuous $O(d,d)$ group of transformations involving the spatial parts of the metric $g_{\mu \nu}$ and  Kalb-Ramond tensor $B_{\mu \nu}$, as well as the dilaton $\phi$. Such a symmetry has been argued to hold only at tree level in the string loop expansion but to all orders in the $\alpha'$ expansion  \cite{Sen:1991zi}, a property that has been explicitly checked up to ${\cal O}((\ap)^3)$ \cite{Meissner:1996sa,Codina1,Codina2}.
	In the particular case in which $B_{\mu \nu}$ is set to zero (to which we limit our considerations in this paper), such a symmetry reduces to a discrete $Z_2^d$ scale-factor-duality (SFD) group together with time reversal $T$ \cite{Veneziano:1991ek,Tseytlin:1991wr}. In the strictly isotropic case the symmetry is  further reduced to the four-dimensional group $Z_2^{SFD} \otimes Z_2^T$.
	
	In a recent impressive paper,  Hohm and Zwiebach (HZ) \cite{HZ}   have shown that, modulo field redefinitions and integrations by parts, the most general reduced (i.e. just time-dependent) cosmological action takes a particularly simple and  manifestly $O(d,d)$-invariant form. For the case of an isotropic FLRW Universe the HZ action depends on just a single even function $F(H)$ of the Hubble parameter $H(t)$. Further developments of the HZ approach have been given in several subsequent papers \cite{21a,3,4,25,26,rost,Bieniek:2022mrv,27,Song:2023txa,Bernardo:2021xtr}.
	
In a very recent paper two of us \cite{Gasperini:2023tus} have reformulated in a Hamiltonian-like formalism (see Sect. 2 for more details) the HZ result, and have shown that this allows to characterise in a simple way the conditions\footnote{These conditions include the requirement that the HZ function $F(H)$,  describing the $\ap$-corrected gravitational Lagrangian density, is non-analytic in the complex-$H$ plane, as expected \cite{Das:1986da} from the known coupling between massless and massive string modes. The importance of a non-trivial branch-point structure in $F(H)$ has also been stressed in a recent application of this formalism to two-dimensional black holes \cite{Codina:2023nwz}.} under which the field equations lead to regular cosmologies smoothly connecting  an initial phase of low-energy pre-big bang evolution to its $SFD \otimes T$-related post-big bang configuration. Under these conditions, the solutions (that can be interpreted as a class of possible ``string vacua") rather than breaking (say in the isotropic case)  the original $Z_2^{SFD} \otimes Z_2^T$ symmetry down to 
nothing (which is the case at lowest order in $\alpha'$, whereby four distinct solutions are generated by the symmetry group \cite{7}), 
are instead invariant under a diagonal $Z_2$ subgroup, connecting just pairs of duality-related  solutions. As a very special case it was easy to recover the regular bouncing solution found  by a trial and error procedure in \cite{Wang1,Wang2}.

Resolving the curvature singularity separating the pre- and post-bounce phases would remove, of course, one of the most important obstacles facing the pre-big bang scenario.
However, as widely discussed for instance in \cite{8,9}, there are obvious theoretical and phenomenological shortcomings with a naive cosmology connecting the two $SFD \otimes T$-related solutions.

 One of these is related to the choice of initial conditions: in a typical pre-bounce solution both curvature and string coupling ($g_s^2 = e^\phi$) grow. Let us recall, for instance, that in a $d$-dimensional isotropic cosmology the initial time evolution of the dilaton and of the Hubble parameter $H$ is given by
\beq
\phi \sim -\left(1 + \sqrt{d}\right) \ln (-t), ~~~~~~~~~~~~~~ H =  \frac{1}{\sqrt{d} \left(-t \right) }~,~~~~~~~~~~~~~~~~~~~~ t \to - \infty \, ,
\label{earlyt}
\eeq
so that both coupling and curvature go to zero in the far past ($t \to - \infty$). This led \cite{Buonanno:1998bi} to invoke a principle of {\it ``Asymptotic Past Triviality"} governing the Universe's initial conditions. 
Indeed, it was  argued in \cite{Buonanno:1998bi} (see also \cite{Feinstein:2000ja}) that  generic (i.e. inhomogeneous and anisotropic) solutions becoming asymptotically trivial in the far past would be affected, as time grows, by gravitational instabilities leading, chaotically, to the formation of different trapped surfaces in different parts of the Universe, with a stochastic distribution of values  for both curvature and string-coupling and a singularity in the future. Inside each trapped surface the geometry would become increasingly ``velocity dominated" (i.e. with sub-dominant spatial gradients) as one approaches the singularity, and akin to the homogeneous solutions we have discussed above for the pre-bounce phase. 

Thus, in different parts of the Universe, different realisations of the pre-big bang initial conditions will be effectively generated with a whole spectrum of small initial curvatures and couplings\footnote{ This is due to the presence of two classical symmetries resulting in as many arbitrary integration constants \cite{Buonanno:1998bi}.}. 
Regions with sufficiently small initial curvature and coupling would then lead to a long phase of dilaton-driven inflation, as described by Eq.~(\ref{earlyt}). 
In other words, the principle of Asymptotic Past Triviality justifies our assumptions on the ``initial" evolution of the homogeneous solutions, while allowing a whole spectrum of initial data (such a principle, by the way, is also needed for phenomenological reasons \cite{gasp}). In this paper we will restrict our attention to cosmologies satisfying this ``Asymptotic Past Triviality" requirement.

Turning now to the late time duality-related solution, however, this would pose serious problems if it could be trusted. In fact, at late times the dual solution to (\ref{earlyt}) is given by:
\beq
\phi \sim \left(\sqrt{d} -1 \right) \ln (t) ,~~~~~~~~~~~~~~ H =  \frac{1}{\sqrt{d} \,t }~,~~~~~~~~~~~~~~~~~~~~ t \to + \infty \, ,
\label{latet}
\eeq
so that the curvature goes to zero also in the far future but the coupling, instead, blows up. That means that, while we could trust the tree-level, low-curvature approximation in the far past, we cannot do the same in the future even if a regular bounce induced by the $\ap$ corrections copes with the high-curvature intermediate phase. Even if the coupling decreases during the short-lived bounce (as it indeed happens in the solutions discussed in \cite{Gasperini:2023tus}), eventually it will increase without limits implying that string loop corrections --and even non-pertubative effects-- will eventually come into play (when exactly depends of course on the initial data).

These corrections break the duality symmetry, making the post-bounce evolution quite unlike its pre-bounce counterpart. Two important consequences come immediately to mind: $i)$ the generation of a non-trivial dilaton potential, and $ii)$  the turning on of cosmological perturbations and particle production. A third one, more typical of string theory, is the possible stabilization of the extra spatial dimensions (in which strings necessarily live) and the isotropisation of our three-dimensional space.

 In this paper we move a first step in the direction of addressing these important problems by considering, on top of the $\ap$ corrections, the  effects of  a non-perturbative dilaton potential $V(\phi)$. We shall discuss, via both numerical and analytical methods, under which conditions these effects result in a stabilisation of the dilaton and in new kinds of interesting  cosmologies at late times.
It should be recalled, in this context, that the unavoidable presence of a non perturbative potential in a scenario with growing dilaton, and its possible stabilisation effects on both the curvature and the string coupling, were  considered also in previous papers (see e.g. \cite{Brustein:1994kw, Kaloper:1995ey,Easther:1995ba,Gasperini:1996in} and references therein). 
In that case, however, the higher order $\ap$ corrections were missing, and the conclusion was that there was no stable fixed point with frozen dilaton towards which a background starting from the string vacuum could be attracted. Here we will show how such a conclusion may change if we work with the HZ-modified string cosmology equations.
 
 The rest of the paper is organized as follows:
 In Sect.~\ref{basicequations} we generalize the approach of \cite{Gasperini:2023tus} by including a non vanishing dilaton potential. We will also make more precise the meaning of the ``Hamiltonian" reformulation introduced in \cite{Gasperini:2023tus} by connecting it to the so-called Routhian approach to dynamical systems in classical mechanics \cite{LL}. In Sect.~\ref{sec3} we shall discuss, for the duality-invariant case with $V(\phi)=0$, some general properties of  regular isotropic bouncing solutions, extending the results of \cite{Gasperini:2023tus} to more complicated evolutions, and arguing that they all belong to the same topological class of background geometries. 
 In Sect.~\ref{sec4}, we will consider  the effects of a dilaton potential in the isotropic case. We first discuss (Subsect.~\ref{sec41})  the case of a potential
with a local minimum  $V_0 =0$ and show that, under suitable initial conditions given in the regime of low-curvature pre-big bang inflation, there are regular solutions which, after the bounce, asymptotically approach a  FLRW attractor of matter-dominated type with a stabilized dilaton. We then consider (Subsect.~\ref{sec42}) the case of $V_0 > 0$,  in which similar initial pre-big bang conditions  lead again to late-time dilaton stabilisation, but with an associated geometry describing a de Sitter inflationary phase  (in both the String and Einstein frames because of dilaton's stabilization).
In Sect.~\ref{sec43} we discuss the range of initial conditions compatible with these final attractors.
In Sect.~\ref{secaniso} we extend our considerations to the anisotropic case showing that late time attractors with a constant dilaton (and both $V_0 = 0$ and $V_0>0$), when they exist, must be isotropic. 
Sect.~\ref{sec6} summarizes our results and offers some concluding remarks. 
Finally, in Appendix~\ref{appA}, we present an example of solution in which the initial perturbative evolution ends up, at late times, in its time-reversed counterpart.

 
\section{Basic equations in a Routhian formalism} 
\label{basicequations}

We shall write our equations in terms of the background fields defined in the so-called String frame (see e.g. \cite{8,9}), in which the $O(d,d)$ symmetry is manifest.  Limiting ourselves to the  gravi-dilaton system with time-dependent field variables $\{\phi, g_{\mu\nu}\}$ and  a $d$-dimensional spatially flat (but possibly anisotropic) spatial metric, we thus set $\phi= \phi(t)$, $g_{00}=N^2(t)$, $g_{ij}= - \da_{ij}  a_i^2(t)$, where $a_i = e^{\b_i}$. 

In such a case, by using the convenient ``shifted dilaton" variable, defined by
\beq
\fb= \phi - \sum_i\b_i,
\label{11}
\eeq
the effective action for the string cosmology equations, including higher-curvature string corrections to all orders in $\ap$, as well as a non-perturbative dilaton potential $V(\phi)$, can be written, slightly extending \cite{HZ}, as:
\bea
&&
S\equiv  \int dt L = -{1\over 2} \int dt \, N e^{-\fb} \left[ N^{-2} \, \fbp^2 + F\left(N^{-1} \dot \b_i\right) + 2 \, (\ap)^{(d-1)/2} V(\phi) \right]\nonumber , 
\noindent \\ &&
F = - N^{-2} \sum \dot \b_i^2 + \mathcal{O}(\alpha')+\dots,
\label{12}
\eea
where the dot denotes time derivatives\footnote{Note that, besides the presence of $V$, there is also an overall factor $(-1/2)$ w.r.t. the action used in Ref. \cite{Gasperini:2023tus}.}. Here $N^{-1} \dot \b_i=H_i$ is the ith-Hubble parameter, and the function $F(H_i)$ in the isotropic case reduces to the function  introduced in \cite{HZ}: it can be written as an infinite (and not necessarily convergent) series of even powers of the Hubble parameter, and includes, in principle, the all-order $\ap$ corrections predicted by a given string model\footnote{These are only known, unfortunately, at a relatively low order. In the spirit of the pre-big bang scenario we have not included a string-scale-size ``cosmological constant" which arises in non critical dimensions (e.g. for $d \ne d_c =9$ for the superstring) by simply assuming that $(d_c-d)$ compact dimensions are flat and frozen at the string length scale.}. Note that, in order to keep the usual dimensions for $V$, and to make it appear in the action with the same dimensions as the kinetic terms, we have multiplied $V$ by $(\ap)^{(d-1)/2}$. In the following, when $\ap$ is not explicitly written, we shall be using units in which $\ap=1$.

By varying the action (\ref{12}) with respect to $N$, $\b_i$ and $\phi$, and defining $f_i= \pa F/\pa H_i$, we obtain the following Euler-Lagrange equations in the cosmic-time gauge $N=1$:
\beq
\fbp^2= F-\sum_if_i \, H_i +2 \, V, ~~~~~~~~~
\dot f_i= f_i \, \fbp +2 {\pa V\over \pa \phi}, ~~~~~~~~~
2 \, \fbpp = - \sum_if_i H_i + 2 {\pa V\over \pa \phi}.
\label{13}
\eeq
To zeroth order in $\ap$ one has $F=- \sum H_i^2$, and  recovers the well known (see e.g. \cite{8,9}) tree-level low-curvature string cosmology equations. 

In such a context, for any given function $F(H_i)$ we have a corresponding scenario of string cosmology evolution. However, as shown in \cite{Gasperini:2023tus}, the models characterized by a regular bouncing transition, and describing a smooth evolution from the pre- to the post-big bang phase, must correspond to non-holomorphic functions $F(H_i)$ that satisfy, on top, quite complicated equations. To select such models it is better to work with the inverse functions $H_i (f_j)$, which gives the Hubble parameters $H_i$ as a power series in $f_j= \pa F/\pa H_j$. 

To this purpose, as shown in \cite{Gasperini:2023tus}, one can conveniently adopt a ``partial Hamiltonian" approach to the action (\ref{12}) (also known as {\it Routhian} approach in a classical mechanics context, see e.g. \cite{LL}), and perform a Legendre transformation on just a subset of the original coordinates $N, \fb, \b_i$, in our case on just the latter $d$ Lagrangian coordinates $\b_i$. Denoting by $\pi_i$ the momentum conjugate to $\b_i$ one defines:
\beq
\pi_i ={\pa L\over \pa \dot \b_i} = -{1\over 2} e^{-\fb} {\pa F\over  \pa H_i} =-{1\over 2} e^{-\fb} f_i  \equiv e^{-\fb} z_i,
\label{14}
\eeq
where we have introduced, for later use, the more useful rescaled momenta $z_i =-f_i/2$. 
The associated Legendre transformation defines the so-called Routhian ${\cal R}(N, \fb, \pi_i)$ 
\bea
{\cal R}(N,\fb, \pi_i) &=& \sum_i \, \pi_i \, \dot \b_i - L  =  N e^{-\fb} \left[  \frac12 N^{-2} \, \fbp^2 + \frac12 \left(F-  \sum_i \dot \b_i \, {\pa F \over \pa \dot \b_i} \right)+  V\left(\fb + \sum_i\b_i\right) \right],
\nonumber \\
 \frac{\pa {\cal R}}{\pa \pi_i}  &=& \dot \b_i  \,,
\label{15}
\eea
where the latter equation is to be used to express $\dot \b_i$ in terms of the $\pi_i$.
Introducing, as in \cite{Gasperini:2023tus}, a reduced ``Hamiltonian" $h(z_i)$, the  above two equations can be rewritten as
\bea
&&
{\cal R}(N,\fb,  \pi_i) = N e^{-\fb} \left[ \frac12 N^{-2} \fbp^2 +  h(z_i)  +  V\left(\fb + \sum_i\b_i\right) \right] ~,~ 
\nonumber \\ &&
h(z_i) \equiv \frac12 \left(F-  \sum_i \dot \b_i \, {\pa F \over \pa \dot \b_i} \right) = \frac12 \sum z_i^2 + \mathcal{O}(\alpha')+\dots~,~ ~~~~~~~~~
 \frac{\pa h}{\pa z_i} = H_i   \,.
\label{15bis}
\eea
Note that the last equation for $H_i = N^{-1} \dot \b_i$ basically inverts the functions $f_i = f_i(H_j)$, giving $H_i= H_i(z_j) = H_i(-\frac12 f_j)$.

Hence, in this new  context, a given model is specified by the choice of $h(z_i)$ and of $V(\phi)$, and equations (\ref{13}) can be rewritten in Routhian language as a combination of  the Euler-Lagrange equations for $N$ and $\fb$ and the Hamilton equations for $\b_i$, $\pi_i$, namely ${\pa {\cal R}}/{\pa \pi_i} =  \dot{\b}_i \, , {\pa {\cal R}}/{\pa \b_i} = - \dot{\pi}_i$. Using (\ref{14}), and after setting at the end $N=1$, these equations can be finally rewritten in terms of $z_i$ as:
\beq
\fbp^2= 2 \, h(z_i)+2V, ~~~~~~~~~
\dot z_i= z_i \, \fbp -{\pa V\over \pa \phi}, ~~~~~~~~~
 \fbpp = \sum_i z_i \, {\pa h\over \pa z_i} +  {\pa V\over \pa \phi}\, ,
\label{16}
\eeq
where, as usual, the first equation (the Hamiltonian constraint) together with the second set imply the last equation. These are the equivalent of equations (\ref{13}) in the Routhian formalism.

Later in the paper we shall deal with cases in which some of the scale factors coincide. In these cases it is more convenient to deal with just the subset of  distinct scale factors. Let us consider, as the simplest example, the fully isotropic case, $\b_i = \b$.  We can easily go over from a given model specified  by the function $F\left(N^{-1} \dot \b_i\right)$ to the isotropic case by defining  $F\left(N^{-1} \dot \b \right) \equiv~ F\left(N^{-1} \dot \b_i = N^{-1} \dot \b \right)$. Note, however, that when going to the Routhian written in terms of the momenta $\pi$ congiugate to $\b$, we get $\pi = \sum_{i=1}^d \pi_i \rightarrow d \pi_1$ (where we just picked one representative $\pi_i$). The relation:
\beq
H = {\pa \mathcal{R}\over \pa \pi}
\label{16bis}
\eeq
remains  of course valid since it is an immediate consequence of the definition of $\mathcal{R}$. However, it is now more convenient (although not necessary) to define
\beq
z \equiv \frac1d  e^{\fb} \pi = -\frac{f}{2d}\,,
\eeq
  so that $z$ and $H$ coincide to leading order in the $\ap$ expansion
Equation (\ref{16bis}) thus becomes:
\beq
h(z)={d\over 2}z^2 + \mathcal{O}(\alpha')+ \dots ~~~; ~~~~
H(z)={1\over d} {\pa h(z)\over \pa z}= z+ \mathcal{O}(\alpha')+ \dots  \, ,
\label{16a}
\eeq
which can be used, if necessary, to reconstruct $h(z)$ from $H(z)$.
For later use let us rewrite the Routhian equations (\ref{16}) in the present case:
\beq
\fbp^2= 2 \, h(z) + 2\, V, ~~~~~~~~~
\dot z= z \, \fbp -{\pa V\over \pa \phi}, \qquad\qquad
 \fbpp =  z \, {\pa h\over \pa z} + {\pa V\over \pa \phi}\,.
\label{17bis}
\eeq
Generalization of the above procedure to the case discussed in Sect.~\ref{secaniso} in which $\b_i = \b$ for $i = 1, \dots, d$ and $\b_i = \tilde{\b}$ for $i = d+1, \dots,d+n$ is straightforward.
The general equations  (\ref{16}), (\ref{17bis}) will be the starting point of our subsequent discussion.


\section{More on regular isotropic bouncing solutions with $V \equiv 0$}
\label{sec3}

In this section we shall recall a few results concerning exact bouncing solutions of Eqs.~(\ref{17bis}) with $\ap$ corrections but without dilaton potential, and describing the smooth evolution of a $d+1$-dimensional isotropic background from the initial regime of Eq.~(\ref{earlyt}) to the final, duality-related post-big bang regime of Eq.~(\ref{latet}).

We will show, in particular, that the curvature bounce illustrated in \cite{Gasperini:2023tus} can be implemented also in a generalised (and highly non-trivial) way like, for instance, through a phase of oscillating background curvature. However, all possible types of bouncing backgrounds obtained in this context, irrespectively of their (possibly) complicated kinematics, always belong to the same topological class in a sense defined below.

Let us first recall that, as discussed in \cite{Gasperini:2023tus} for the $V=0$ case, the possible existence  of  regular bouncing solutions is controlled by the analytic properties of the function $h(z)$ and its associated ``companion" in Eq.~(\ref{16a}), $H(z)=-H(-z)= (1/d)(\pa h/\pa z)$. For a regular bounce to occur we must require that $h(z)$, which grows from zero to positive values for $z \ll 1$, exhibits a second zero at $z=z_2$ (and, of course, also at $z=-z_2$). Assuming $h(z)$ to be continuous and differentiable, this implies that $H(z)$ itself vanishes at (at least) one point $z_0 < z_2$ and to have local extrema at various points $(z_1, \dots)$ in that interval.

\begin{figure}[ht!]
\centering
\includegraphics[scale=0.55]{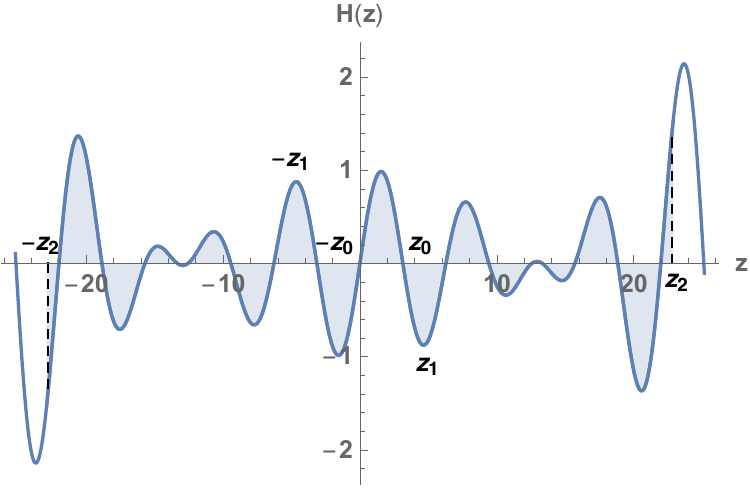} ~~~~~~
\includegraphics[scale=0.55]{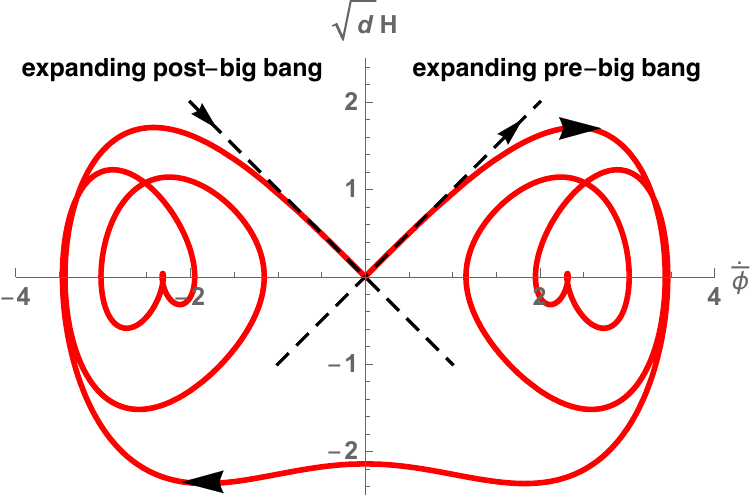}
\caption{Left: the blue curve is the function $H(z)$ of Eq.~\eqref{eq:Hf} for $d=3$ and $\epsilon=-0.5$. Along a given branch, positive or negative, the total shaded area subtended by the curve vanishes. Right: the red curve describes the parametric plot of a numerical solution of Eqs.~(\ref{17bis}) for the previous $H(z)$ and for $V(\phi)=0$. The black dashed bisecting lines 
(satsfying the condition $\fbp= \pm \sqrt{d}H$) represent the asymptotic regimes of initial low-energy expansion from the string vacuum, and  final, post-big bang, decelerated expansion.}
\label{fig:HNegEps}
\end{figure}

Simple examples satisfying such conditions (like, for instance, $h(z) \sim (1/2) z^2\left[1-z^2/2\right]$) have been considered and discussed in \cite{Gasperini:2023tus}. 
There is also, however, the possibility of more complicated models\footnote{Some of these models can be excluded a priori by being impossible to realise in string theory. These includes models in which, at intermediate times or near the bounce, the solution enters the perturbative region in a way incompatible with the known perturbative effective action.} described by a function $H(z)$ which has several extrema (or zeros) in the range $\{0, |z_2|\}$. Consider, for instance, the model described by the following effective Hamiltonian:
\begin{equation}
\frac{h(z)}{d} = 1- \cos z
- \frac{2 \,\ep}{9 \, \pi^2}\left[1+\left({z^2\over 2}-1\right) \cos z - z \sin z \right], 
\label{eq:HamEps}
\end{equation}
which gives, via (\ref{16a})
\begin{equation}
H(z)=\sin\ z \left[ 1+\epsilon\left( \frac{z}{3\pi } \right)^2 \right]\,.
\label{eq:Hf}
\end{equation}
By assuming $\ep<0$, and using the constraint $\fbp = \pm \sqrt{2h}$ following from Eqs.~(\ref{17bis}) without dilaton potential, we can then obtain the corresponding model of regular bounce illustrated in the plane $\{\fbp(z), \sqrt{d} H(z)\}$ by the parametric plot of Fig.~\ref{fig:HNegEps}.

The left panel of Fig.~\ref{fig:HNegEps} is the plot of the curve $H(z)$. The zeros of $H$ correspond to the values of $z$ at which the curve intersects the horizontal axis, such as $\pm z_0$ positions of the points $z_0$, the extrema of $H$ to maxima and minima such as $\pm z_1$. The shaded area subtended by the curve, for both the positive and the negative range of $z$, vanishes, such that the points $\pm z_2$ marking the end of the shaded region correspond to the intersection of the curve with the vertical axis. Also, the contribution of $\ep$
introduces a modulation of the peaks which leads to a series of alternated local minima and maxima. 

This last property is present also in the parametric plot of the right panel, producing a series of points where $|H|$ reaches a local maximum or a zero (corresponding to a transition from expansion to contraction or viceversa), even for $\fbp$ non-vanishing. The existence of these peculiar points leads to the ``vortex-like" red curve of Fig.~\ref{fig:HNegEps}. Note that the 
turning points of the parametric curves around $H=0$ and $|\dot{\bar\phi}|\simeq 2.25$ are smooth, even though they might look cusps: this is only due to the overall size of the plot. It should be stressed, finally, that for $\ep \ra 0$ the ``bottom" of the red curve,  corresponding to the point $\fbp=0$
which formally marks the transition from the pre- to the post-big bang regime (i.e. from the right to the left sector of the parametric plane), tends to approximate the origin from below, $H \ra 0_-$. In that case one would just recover an example of the cases mentioned in the previous footnote 6, as the solution would follow a low energy trajectory incompatible with the perturbative string cosmology equations.

The given example clearly displays the possibility of regular bouncing scenarios described by solutions which are always self-dual in the absence of a dilaton potential, but which may be characterised by a high-energy string phase with rapidly oscillating Hubble parameter, implying sudden (but smooth) transitions connecting expanding $\leftrightarrow$ contracting geometries.

Hence, in this general context, the word ``bounce" in no way should be interpreted as meaning a localised transition from initial contraction to final expansion. More appropriately, with the word ``bounce" we mainly refer in this paper to the absolute value of the curvature scale, occurring during a possibly extended (in string units) epoch needed to convert the initially accelerated, growing curvature expansion to the final decelerated, decreasing curvature expansion. 

The class of solutions compatible with this scenario in principle is large, and controlled by the analytical properties of $h(z)$ (whose correct expression should be provided by string theory). However (in the absence of a dilaton potential), all such solutions are topologically equivalent with respect to the following geometric property of the curve describing the given solution in the parametric plot of Fig.~\ref{fig:HNegEps}: the vector connecting the origin to a point on the curve undergoes a clockwise rotation of $3\pi/2$ as one goes from the beginning to the end of the curve. The same is true for  solutions smoothly connecting a contracting initial phase to the related final contracting one. The only difference is that the initial and final parts of the curve lay in the bottom-left and bottom-right sectors of the plane, respectively, and thus (as already noted in \cite{Gasperini:2023tus}) the rotation occurs in the anticlockwise direction.


\section{Regular isotropic bouncing solutions and dilaton stabilisation with $V\ne 0$}
\label{sec4}

As discussed in the introduction a tree-level string cosmology, even if it implements a regular bounce, cannot be realistic. As a first step towards making the model more realistic let us now include into the effective string cosmology equations the contributions of a non-perturbative dilaton potential $V(\phi)$, which goes to zero in the small coupling limit $g_s^2 \ra 0$ ($t \ra -\infty$) with an instanton-like suppression of the type $V \sim e^{-{\rm const}/g_s^2}$, and which becomes non-negligible in the opposite large time limit, thus breaking the duality symmetry and modifying the final, post-bounce asymptotic configuration. We may expect, in this way, not only a modified dilaton dynamics but also a modified final evolution of the cosmic geometry  (no longer necessarily duality-related to that of the initial low-energy solution).

We are interested, in particular, in a ``realistic" post-bounce scenario with the dilaton stabilized at a final constant value $\phi_0$ such that $g_s^2(\phi_0) = e^{\phi_0} \laq 1$, and in which the asymptotic solution can approach the phase of standard cosmological evolution described by the Einstein gravitational dynamics (with no need of string loops and/or $\ap$ corrections). We shall thus consider an effective dilaton potential which has a local minimum $V=V_0$, needed to stabilise the dilaton, and which can be parametrised in a phenomenological way (and in units $\alpha'=1$) as follows: 
\beq
V(\phi)= A \, e^{-{B(\phi)/ \b}}\left[\left(c^2- B(\phi)\right)^2 + \da B(\phi) \right]
\left[1 - q \, B^{-1}(\phi) \right], 
\label{41}
\eeq
where
\beq
B(\phi)= {1+\a \, g_s^2 \over \a \,  g_s^2}= {1 + \a \, e^\phi\over \a \, e^\phi},
\label{41a}
\eeq
and where $A$, $c$, $\a$, $\b$, $\da$ and $q$ are constant parameters controlling various features of $V$.

In particular, $A$ (together with $c$) controls the overall magnitude of the potential; $B$ is some kind of inverse 't Hooft coupling, $\lambda_t^{-1}$, in the weak coupling limit (with $\alpha \sim N_c$, the number of colors) while it approaches from above a finite value, here conventionally set to one, in the strong-bare-coupling limit\footnote{This is the idea of the dilaton runaway scenario \cite{Piazza} (see also \cite{Damour:2002mi,Damour:2002nv} for its possible observable consequences) which is based on the assumption \cite{Veneziano:2001ah} that the limit $\phi \to + \infty$ is non singular and characterized by finite (and perhaps realistic)  values for both the gauge and gravitational coupling (in string units). However,
our simple description of that regime (through the relation between $B$ and $\phi$) is over-simplistic since it ignores the fact that loop corrections, besides generating a non trivial potential, will also modify the whole kinetic part of the action \cite{Piazza}. This, in turn, would make the passage from the S-frame to the E-frame more complicated than in the perturbative regime.}. 

\begin{figure}[t]
\centering
\includegraphics[width=7.5 cm]{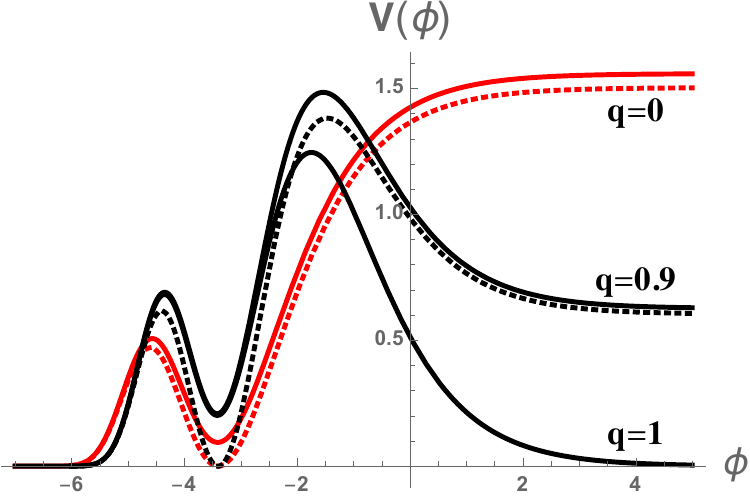}
\caption{The red curves show an example of dilaton potential 
asymptotically approaching the maximum at large $\phi$ values, and obtained from (\ref{41}) with $q=0$. The black curves show an example of ``runaway" potential, asymptotically going to zero for $q=1$ and stabilising to a non-vanishing constant value for $0<q<1$. 
Solid curves are plotted with $\da >0$, and are characterised by a  local minimum $V_0>0$. Dotted curves correspond to the same potential plotted however for $\da =0$, and with a local minimum $V_0=0$. All curves are plotted for $c=2$, $\a=10$. We have used $A=0.22$, $\b=3.5$ for the red curves, and $A=1$, $\b=2.5$ for the black curves.}
\label{f40}
\end{figure}

In the presence of a local minimum of $V$ at $\phi=\phi_m$, the parameter $\delta$ controls  $V_0\equiv V(\phi_m)$ i.e.  $V_0$ is non-vanishing if and only if $\da$ is non-vanishing  (as illustrated in Fig.~\ref{f40}). 
The position $\phi_m$ of the local minimum, if present, 
is mainly controlled by $c$ and $\a$, while the parameter $\b$ mainly controls the height of the first potential peak.
Also, the asymptotic behaviour of $V(\phi)$ at large positive values of $\phi$ depends on the parameter $q$: for $q\leq 0$ the potential approaches there its maximal constant value; for $0<q<1$ the potential approaches from above a non-vanishing positive constant\footnote{Such a constant is negative if $q>1$, but since in this paper we are mainly interested
in a scenario where the dilaton growth is trapped by the potential (see below) without reaching the large $\phi$ regime, it will be enough to concentrate our discussion on the range $0\leq q \leq 1$.}. Finally, for $q=1$ the potential is asymptotically vanishing thereby realizing the so-called ``dilaton runaway scenario" of \cite{Piazza} (see also \cite{Damour:2002mi,Damour:2002nv}).
Fig.~\ref{f40} gives a simple qualitative illustration of the various possible cases.

The specific ``shape" and the amplitude of the potential 
strongly depend on the numerical values of the various phenomenological parameters, and the values used in Fig.~\ref{f40} have been chosen mainly for the purpose of a clear graphical illustration of the possible differences.  What is important to stress is that, depending on the values of such parameters (and on the initial conditions which identify the particular background whose evolution we are following from the asymptotic, low-energy regime), the potential may significantly affect the background evolution not only during the bounce, but also, and most important, in the final asymptotic post-bouncing regime. In addition, as we shall see below, the dilaton's evolution is best understood in the E-frame, in which it behaves like a minimally coupled scalar, while the potential in (\ref{41}) refers to the S-frame. We stress again that the well known connection between the two frames in the perturbative regime can be strongly modified at large positive $\phi$ in the runaway scenario of \cite{Piazza}. This being said we may expect, in general, three possible late-time dilaton evolutions. 

The first  case is the one in which the potential is unable to substantially  modify the overall dilaton evolution: the dilaton keeps monotonically growing both before and after the bounce, and (with the appropriate model of $\ap$ corrections) we recover a regular  transition from the expanding pre- to post-bang regime like in the cases with no potential (see \cite{Gasperini:2023tus} and Sect.~\ref{sec3}). This will typically happen if the overall scale $A$ of the potential is small enough as compared to the value of $H$ and $\dot{\phi}$ when the dilaton is in the region of sizeable potential. Describing the late time physics corresponding to that case is not simple and will depend on whether the Einstein-frame potential falls asymptotically to zero, to a positive constant, or even grows indefinitely. Since the passage to the E-frame is not simple at large bare coupling, we shall postpone this case to some future work.

The second case is the one in which the potential is high enough to stop  the growth of dilaton,  and the dilaton bounces back towards the small coupling regime, monotonically approaching the asymptotic limit $\phi \ra -\infty$: in that case the final background configuration after the bounce is exactly the time-reversed of the initial one, thus implementing a (new type of) regular bounce from expanding pre- to contracting post-bang regimes (see Appendix~\ref{appA}). 

Finally, the third (and phenomenologically most interesting) case is the one in which the dilaton gets trapped in the local minimum of the potential. The rest of this paper will be devoted to illustrate and discuss this last possibility, which has three important consequences. With an appropriate choice of the parameters of Eq.~(\ref{41}),  the potential can produce $i)$ the stabilisation of the dilaton at a final asymptotic value $\phi=\phi_0=$ const; $ii)$ a final evolution of the metric of standard type, corresponding to a dust-dominated FLRW geometry if $V(\phi_0)=0$ or to a de Sitter geometry if $V(\phi_0)>0$; $iii)$ the isotropisation of the final geometry if we start from anisotropic initial conditions.
The first and second effects will be studied in this section, where we will concentrate on the case of an isotropic scenario. The isotropisation phenomenon will be discussed in  Sect.~\ref{secaniso}.


\subsection{FLRW attractors for a local minimum $V_0 = 0$}
\label{sec41}

For a first illustration of the dilaton stabilisation mechanism
we will start considering an initially expanding $(d+1)$-dimensional isotropic background geometry, asymptotically evolving from the string perturbative vacuum according to Eqs.~(\ref{16}), and a dilaton potential given by Eq.~(\ref{41}) with $\da=0$ (such that $V=0$ at the local minimum  $\phi=\phi_m$, see the dotted curves of Fig.~\ref{f40}).

Also, to stress the differences induced by the potential on the  evolution of the background geometry, let us directly present a numerical integration for the same model of $\ap$ corrections producing the regular bounce first derived in \cite{Wang1}. As shown in \cite{Gasperini:2023tus}, such a model provides a regular solution of the Lagrangian equations (\ref{13}) corresponding to the inverse  HZ function $H(f)$ given by
\beq
H(f)=-{f\over 2 \, d}+\ap \left(f\over 2 \, d\right)^3.
\label{42}
\eeq
Adopting the Hamiltonian formalism for isotropic backgrounds, and using in particular Eqs.~(\ref{16a}) (recalling that in the isotropic case, $z=-(f/2d)$),  we obtain that the above model is described by the effective Hamiltonian\footnote{The expression given in Eq.~(\ref{43}) matches, up to the first $\ap$ correction, the expression of $F(H)$ for the heterotic string \cite{Codina2}.}
\beq
h(z)= {d\over 2} \left( z^2 - \ap {z^4\over 2}\right) \, ,
\label{43}
\eeq
where, unlike in the previous paper  \cite{Gasperini:2023tus}, we will use units in which $\ap=1$. 

Starting with this result, we can now easily provide a qualitative illustration of the dilaton stabilisation (and of the related effects)  produced by the potential by performing a numerical integration of Eqs.~(\ref{17bis}), for any given model of potential specified by the parameters of Eq.~(\ref{41}). We shall consider, to this purpose, 
a simple example of runaway potential with $q=1$, but the results we are presenting can be reproduced for other classes of potentials with $0\leq q<1$, and with different asymptotic behaviour (see Fig.~\ref{f40}).  It should be stressed, however, that the  value of $q$ (and of other  parameters) may be relevant to define the region of initial conditions compatible with the attraction to the final regime with stabilised dilaton, as we shall discuss in Sect.~\ref{sec43}.

\begin{figure}[t]
\centering
\includegraphics[width=5.5cm]{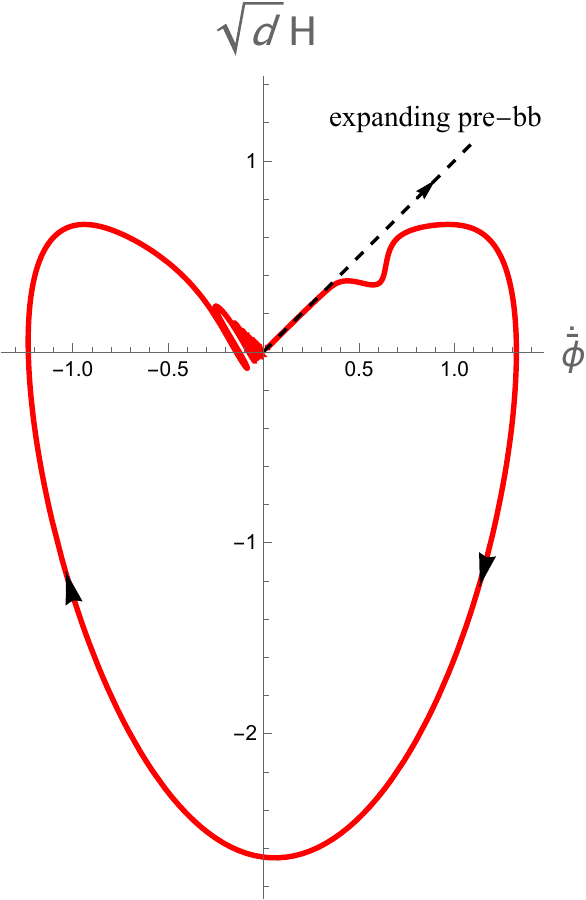}
\caption{The red curve describes the parametric plot of a numerical solution of Eq.~(\ref{16}), with the potential (\ref{41}) and the Hamiltonian (\ref{43}). We have set $d=3$ and $\ap=1$ for the Hamiltonian, $A=0.1$, $\a=10$, $\b=3$, $c=2$, $\da=0$, $q=1$ for the potential, and $\phi=-3.5$, $z=0.01$ for the initial conditions. 
The black dashed half-line corresponds to the initial trajectory evolving from the string perturbative vacuum and described by Eq.~(\ref{earlyt}).}
\label{f41}
\end{figure}

We shall impose on our initial conditions, fixed in the region where the dilaton potential (and the $\ap$ corrections) are still negligible, to satisfy the low energy pre-big bang equations $\fbp = \sqrt{d} H$.  
The numerical solution we obtain, with appropriate (small enough) values of the initial condition for $\phi$ and $z$, gives then the results illustrated by the parametric plot of Fig.~\ref{f41}. 
As already stressed we have chosen a potential with $\da=0$ and $q=1$, and the numerical values of the other parameters are specified in the caption of the figure.

Note that we find again a regular bounce, described by a smooth curve turning clockwise in the plane of the figure (as repeatedly stressed in \cite{Gasperini:2023tus}). With respect to the case without potential, however, the curve describes a ``deformed heart-like" path (no longer symmetric with respect to the vertical axes), and we have two types of deformations. A first deformation occurs in the pre-bounce regime (the upper right quadrant of the figure), where the effects of the potential first come into play (together with those of the $\ap$ corrections). The physically more significant deformation occurs however in the final, post-bounce regime (upper left quadrant), corresponding to a drastic change of the dilaton dynamics because of its trapping in the potential minimum. The produced result is an oscillating final regime, as shown in Fig.~\ref{f41}.

\begin{figure}[t]
\centering
\includegraphics[width=8.5cm]{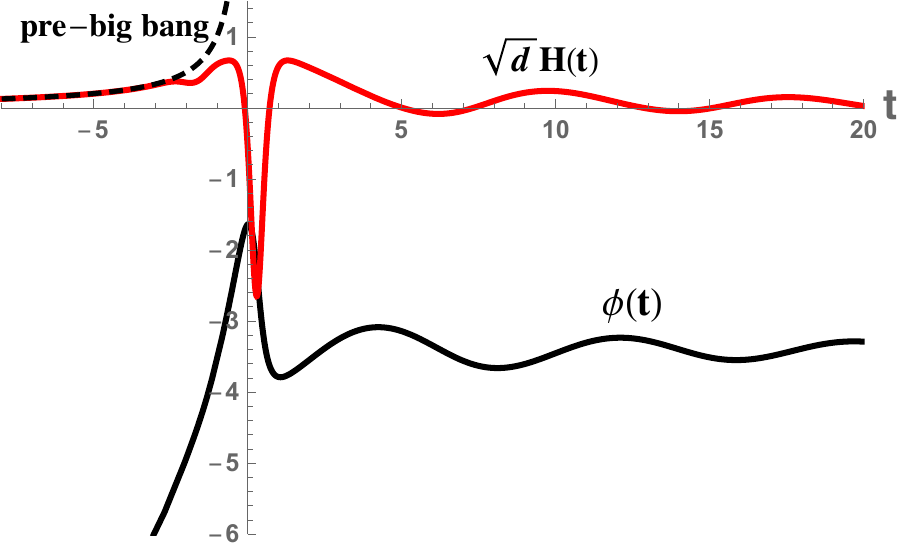}
\caption{Time evolution of $H$ (red curve) and $\phi$ (black curve) for the same numerical solution of Fig.~\ref{f41}. The black dashed curve describes the (unbounded) growth of the Hubble parameter for the low-energy pre-big bang solution of Eq.~(\ref{earlyt}).}
\label{f42}
\end{figure}

The final asymptotic effect of dilaton stabilisation and background oscillations can be explicitly illustrated also by plotting the time behaviour of the numerical solution for $H(t)$ and $\phi(t)$. The result, shown in Fig.~\ref{f42}, again emphasises the differences between the initial and final regimes which (because of the potential) are no longer duality related. 

\begin{figure}[t]
\centering
\includegraphics[width=7.5cm]{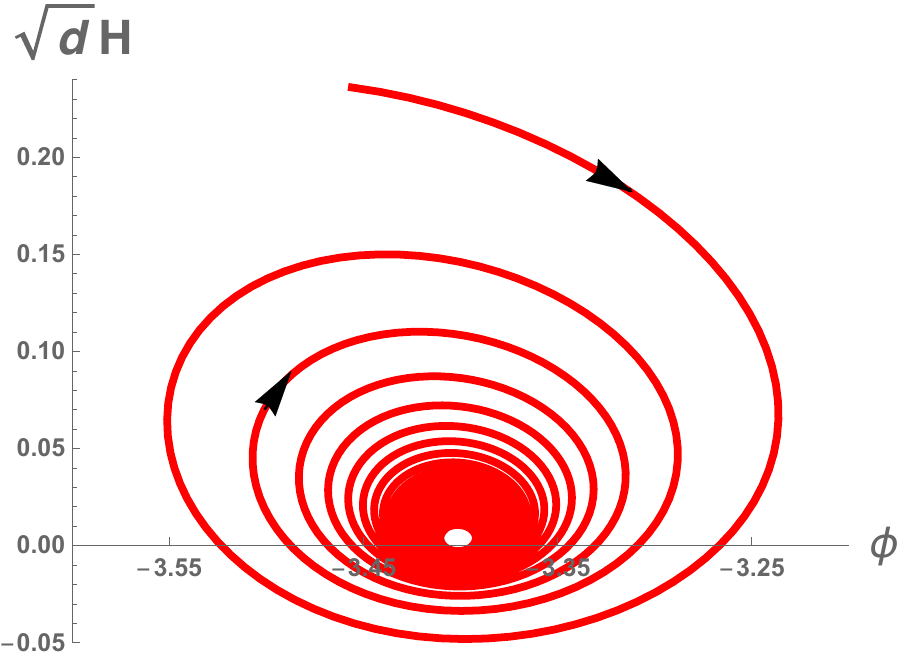}~~~~~~~~~~~~~~~~~~
\includegraphics[width=4.5cm]{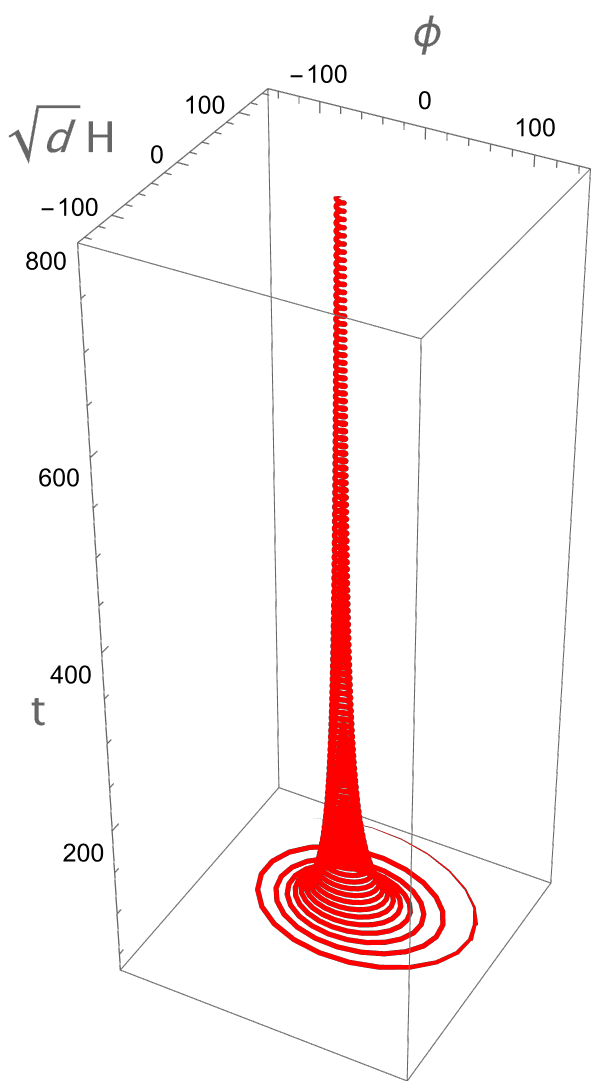}
\caption{Left: two dimensional parametric plot of $H=H(\phi)$. Right: its three-dimensional version with explicit time evolution, $H(t)=H[\phi(t)]$, for the numerical solution of Fig.~\ref{f41}. In the three-dimensional plot the variables $H$ and $\phi$ have been multiplied by $10^3$ for a better graphic illustration of the damped oscillations regime.}
\label{f43}
\end{figure}

The dilaton stabilisation and oscillation effects are even more evident if we consider the parametric plot of $H= H(\phi)$, or its $3d$-version (i.e. the same parametric plot expanded as a function of time). Limiting our attention to the final, post-bounce regime we obtain the results shown, respectively, by the left and right sectors of  Fig.~\ref{f43}. They clearly show that the oscillating background approaches an oscillating regime where the dilaton asymptotically reaches a final (constant, non vanishing) value $\phi_0<0$, and the Hubble parameter is asymptotically decreasing to zero. 

To obtain a more precise information on the type of final geometry and show, in particular, that it describes a phase of standard (FLRW type) evolution (and not a post-big bang evolution of the string cosmology type), it is convenient to discuss the analytical solutions of Eq.~(\ref{17bis}) in the late-time regime where the final asymptotic value of the dilaton $\phi_0$ coincides with the position of the local minimum $\phi_0=\phi_m$, the trapped dilaton is oscillating around $\phi_m$, and the potential can be approximated in functional form as:
\beq
V(\phi)\simeq {1\over 2} m^2 \left (\phi- \phi_m\right)^2.
\label{44}
\eeq
If the potential (\ref{41}) has a local minimum, and if $\da=0$, then the minimum $V_0=0$ is located at $\phi_m= -\ln[\a(c^2-1)]$. In such a case, by expanding $V(\phi)$ up to second order around $\phi_m$, we can also find the value of $m^2$ in terms of the other parameters, and we obtain   $m^2=(2A/c^2)(c^2-q)(c^2-1)^2 \exp(-c^2/\b)$. 

We also note that, at late enough times (depending on the given initial conditions), the curvature scale is small enough (in units $\ap$) so that the higher order $\ap$ corrections can be neglected. With these approximations, the equations (\ref{17bis}) can be rewritten as 
\bea
&&\left( \dot \phi - d \, H \right)^2 = d \, H^2 + m^2 \left( \phi - \phi_m \right)^2,
\nonumber \\ &&
\dot H = H \left( \dot \phi - d \, H \right)- m^2 \left( \phi - \phi_m \right),
\nonumber \\ &&
\fpp= d \, \dot H +d \, H^2 +m^2 \left( \phi - \phi_m \right).
\label{45}
\eea
Looking for oscillating solutions, compatible with the asymptotic limit $\phi= \phi_0=\phi_m$ and $H=0$ for $t \ra + \infty$, we set:

\beq
\phi(t)- \phi_0\simeq {M\over t^\ga} \sin (\om t + \theta)
\label{46}
\eeq
(where $\theta$ is an arbitrary phase parameter), and we find that the above equations (\ref{45}) are satisfied, to leading order in $1/t^\ga$, provided that $\ga=1$, $\om=m \sqrt{d-1}$, and 
\beq 
H(t)= {1\over t} {M \om\over \sqrt{d} \, (d-1)} \left[1 + \sqrt{d} \, \cos (\om t + \theta) \right].
\label{47}
\eeq
Finally, by expanding our solution for small values of $M$ and $\dot M/M\om$, we find at the first sub-leading order that $ M \om \simeq 2 (d-1)/\sqrt{d}$, and that our asymptotic solution can be written as:
\bea
&&
\phi(t)= \phi_0+ {2 (d-1)\over t\,\om \sqrt{d}}  \sin (\om t + \theta),
\nonumber \\ &&
H(t)= {2\over t\,d} \left[1 + \sqrt{d} \cos (\om t + \theta) \right].
\label{48}
\eea
We have checked that the percent difference between these analytical expressions and the previously presented numerical solutions is very small ($\laq 10^{-7}$) at large times.

Note that the time-averaged  behaviour of the geometry gives $\left < H \right >\simeq (2/d)t^{-1}$, corresponding to a scale factor $a(t) \sim t^{2/d}$, which exactly reproduces the time evolution of a standard, dust-dominated, FLRW cosmology. The role of the effective dust fluid, in this case, is played by the oscillating dilaton, which produces a phase of final post-bounce evolution very similar to the dust-like phase of post-big bang evolution dominated by the oscillations of the Kalb-Ramond axion (see e.g. \cite{Bozza1,Bozza2,Gasperini:2003pb}).

What may look surprising, however, is that the oscillations in (\ref{48})  have a large and  constant amplitude (relative to the non-oscillating term) forcing $H(t)$ take both positive and negative values\footnote{We are grateful to Robert Brandenberger and Jerome Quintin for having raised this point.}. This is due to the fact that, as already mentioned, the dilaton does not behave in the string frame  as a minimally coupled scalar. On the other hand, we might have expected that the stabilisation of the dilaton leads to the asymptotic identification of the S and E-frames.

The computation of the E-frame Hubble parameter (see e.g. \cite{9}) gives, from Eq.~(\ref{48}) (and in terms of the S-frame cosmic time $t$):
\beq H_E(t)= e^{\phi/(d-1)} \left ( H - {\dot \phi\over d-1} \right) \equiv 
{2\over t\, d} \,{e^{\phi/(d-1)}} \, .
\label{49}
\eeq
We find again the dust-like behaviour, $H_E \sim (2/d)t^{-1}$, but this time with very small and damped oscillations due to the presence of $\phi(t)$ in the exponential factor. For $\phi\rightarrow \text{const}$, the E-frame time coordinate is simply given by $t_E=t \exp[-\phi/(d-1)]$, and one exactly recovers dust dominated evolution $H_E=2/(t_E\,d)$. In other words, even if the dilaton has damped oscillations, the passage to the Einstein frame is important for it to behave like a genuine minimally-coupled massive scalar. 

Let us finally discuss, for the above given potential, the  range of initial conditions compatible with the scenario we have discussed, i.e. the trapping of the dilaton in the $V_0=0$ minimum and the final associated regime of standard (dust-like) cosmic evolution. In particular, we would like to find out whether or not such initial conditions are highly fine-tuned.

In the isotropic case (see Sect.~\ref{secaniso} for the extension to anisotropic cosmologies)  the initial conditions consist of giving  $\phi$, $\dot{\phi}$ and $H$ at some initial time $t_0$ in the far past. However, for any given model  (i.e. for a given $h$ and $V$) the Hamiltonian constraint (i.e. the first of Eqs.~(\ref{17bis})) can be used to fix the initial value of $\dot{\phi}$ in terms of the other two. Furthermore, assuming that the evolution starts at sufficiently small coupling $e^{\phi}$, where $V$ is absolutely negligible, the second of Eqs.~(\ref{17bis}) implies the conservation law $e^{-\fb} z = \kappa^{-1}$, namely:
\beq
e^{\phi} = \kappa \, z (a^d)  \sim \kappa z (\sqrt{d} \, H)^{\sqrt{d}}  \sim \kappa \, (\sqrt{d} \, H)^{1+ \sqrt{d}} \sim \kappa \, (\sqrt{d} \, z)^{1+ \sqrt{d}}  \, ,
\label{49bis}
\eeq
with $\kappa$ a constant  and the various approximate relations becoming exact in the $t \to - \infty$ limit. This means that changing $\kappa$ does change the initial conditions.  Instead, changing $\phi$ and $H$ (or $z$) while keeping  $\kappa$ fixed, amounts physically to the same initial conditions simply referred to a shifted initial time. In other words, we expect the basin of attraction to the FLRW late time solution to be one-dimensional. It is given by some interval(s) in $\kappa$, a physical quantity given by a suitable combination of the coupling constant and curvature that remains constant during the very early time evolution.
See in particular sect.~\ref{sec43} for a discussion of the basin of attraction towards a final regime of  stabilised dilaton, decreasing curvature and decelerated expansion  illustrated in this subsection.


\subsection{de Sitter attractors for a local minimum $V_0 > 0$}
\label{sec42}

Another interesting cosmological scenario, still describing a regular bouncing evolution from the string perturbative vacuum but approaching a different final configuration, can be obtained with a slight modification of the dilaton potential used in the previous section.

Starting again with the general form (\ref{41}) of $V(\phi)$, but including also the contribution of a (small) non-vanishing parameter $\da$, one finds that, in the presence of a local minimum, the latter is still approximately  located for small $\delta$ at $\phi=\phi_m \simeq - \ln [\a (c^2-1)]$ (as illustrated in Fig.~\ref{f40}), but the associated potential energy is in general non-vanishing. In the limit $\da \ll1$, by computing the potential at   the above minimum to first order in $\da$ , we obtain  in particular $V_0= V(\phi_m)\simeq A \da  (c^2-q) \exp(-c^2/\b)$.

Such a difference is important because the local minimum responsible for the dilaton stabilisation  mechanism
also controls the asymptotic value of the post-bounce Hubble parameter, and for $\da>0$, $V_0>0$ one finds that the background geometry approaches a final phase of standard de Sitter evolution. 
In that case, as we shall see, the final value $\phi_0$ of the dilaton no longer coincides with the position of the local 
minimum, i.e. $\phi_0\not=\phi_m$.

 In order to illustrate the qualitative aspects of this modified scenario  it may be appropriate to present, first of all, the results of a numerical integration of Eqs.~(\ref{17bis}) using for $h(z)$ exactly the same model of $\ap$ corrections as in the previous section~\ref{sec41}, and specified by Eq.~(\ref{43}). In such a way all differences are only due to the modified potential, and in particular to the new contribution of a parameter $\da \not=0$. 
 
\begin{figure}[t]
\centering
\includegraphics[width=5.5cm]{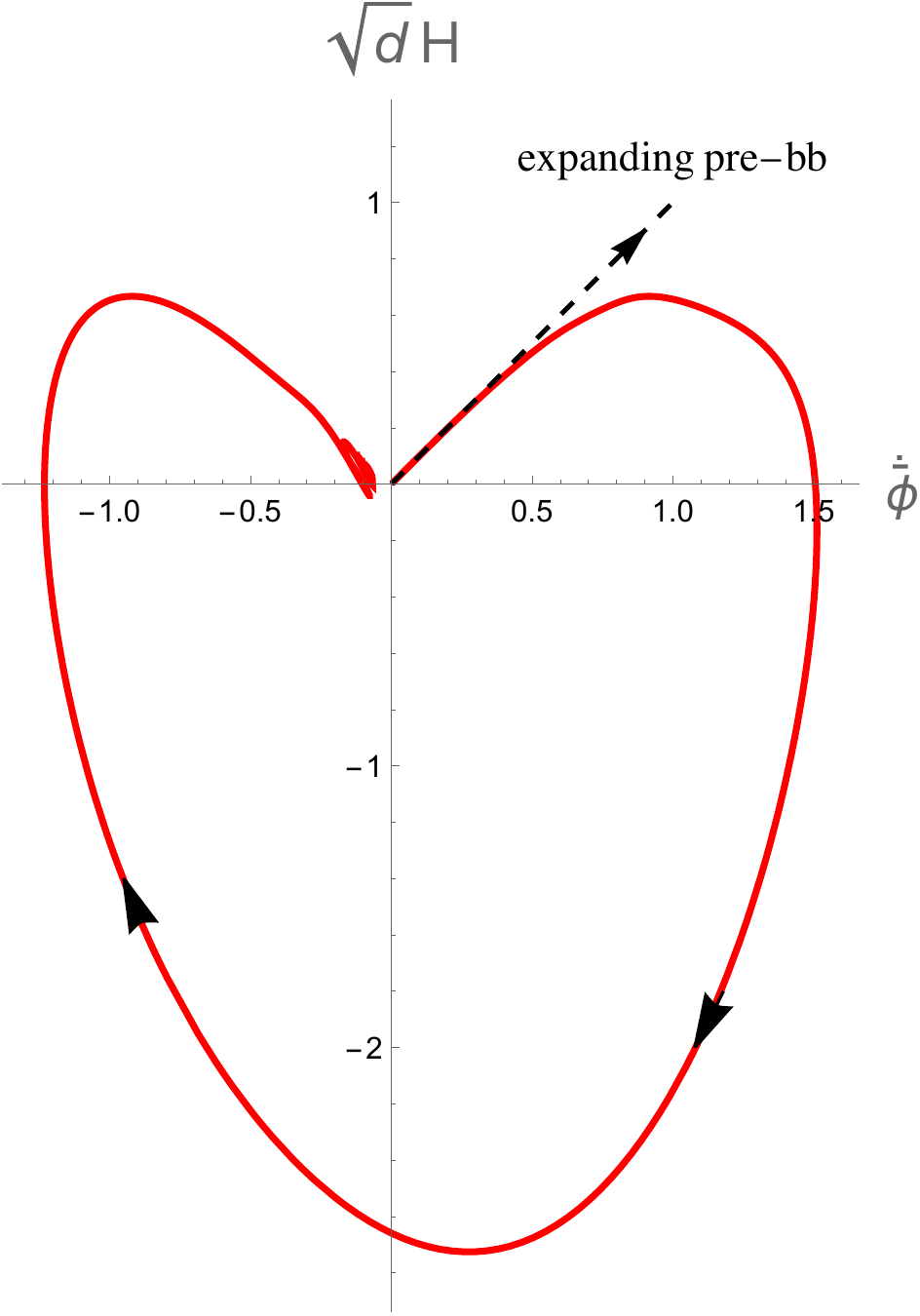}~~~~~~~~~~~~~
\includegraphics[width=8 cm]{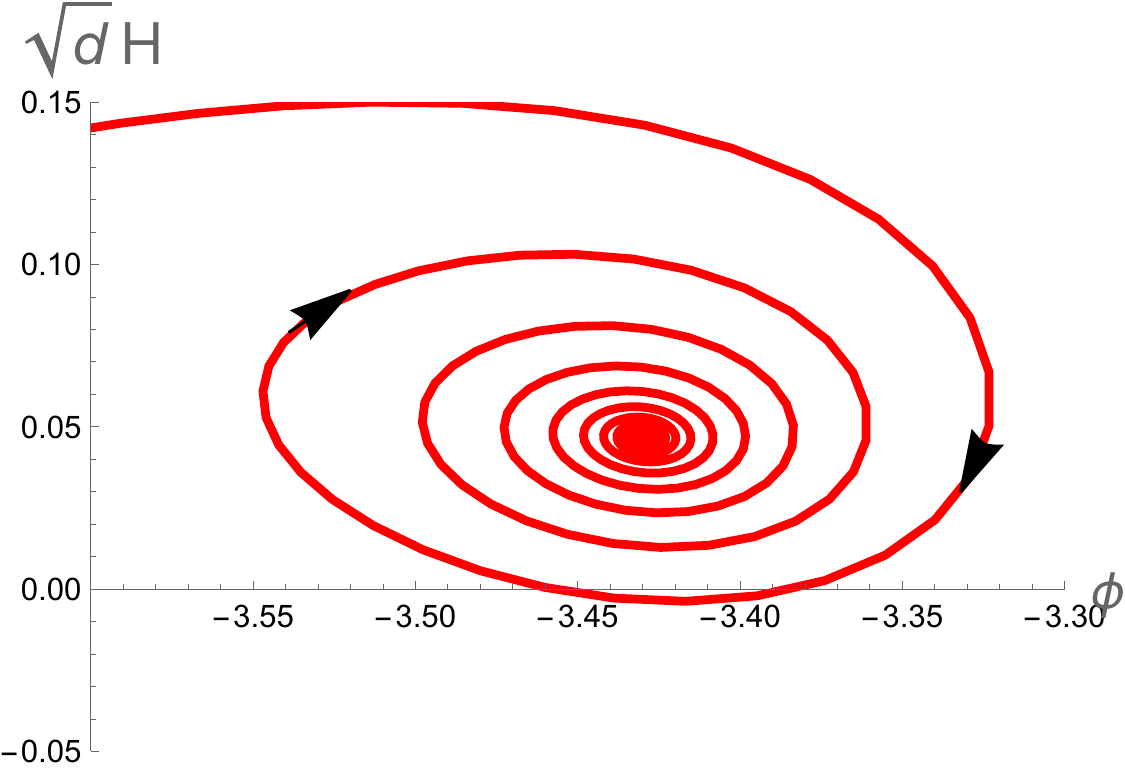}
\caption{Left: the red curve describes the parametric plot of a numerical solution of Eq.~(\ref{17bis}), with the potential (\ref{41}) and the Hamiltonian (\ref{43}) (with $\ap=1$ and $d=3$). Differently from the previous figures we have set $\da=0.04\not=0$ to have a non-vanishing $V_0$ at the local minimum, and $\b=1$, $A=1$ (to simplify the graphics). All the other parameters are the same as in the figures of Sect.~\ref{sec41}: $\a=10$, $c=2$, $q=1$, and the initial conditions are $z=0.01$ and $\phi=-3.65$. The black dashed half line describes the initial trajectory evolving from the string perturbative vacuum, as in Fig.~\ref{f41}. Right: Parametric plot $H=H(\phi)$ for the same numerical solution, in the asymptotic range of large positive times.}
\label{f44}
\end{figure}

We shall impose initial conditions satisfying as before the low-energy pre-big bang dynamics, specified (for $t \ra - \infty$) by the constraint $\fbp= \sqrt{d} H$. The numerical integration of Eqs.~(\ref{17bis}) with the particular value $\da=0.04$ then gives the result illustrated in Fig.~\ref{f44}. The left sector shows again a regular curvature bounce from the pre- to  the post-big bang regime, similar to the case considered in the previous section and illustrated in Fig.~\ref{f41}. The differences from that case are all concentrated in the final post-bounce regime (upper-left quadrant), showing that now the background oscillations due to the stabilisation of the trapped dilaton are no longer approaching the origin of the axes $\{\fp=0, H=0\}$, but a new asymptotic (de Sitter) limit with $H=H_0=$ const, and $H_0>0$.

These oscillations around the new final attractor $\{\phi_0, H_0\}$ are more clearly shown if we concentrate our graphic analysis on the post-bounce region of large positive times, as illustrated by the parametric plot $H=H(\phi)$ presented in the right sector of Fig.~\ref{f44}, where we have used the previous numerical solution in the range $t>0$. The final value of $H_0$ is controlled by the value of the parameter $\da$ in the dilaton potential, and for $\da \ra 0$ one recovers the FLRW scenario with $H \ra 0$ discussed in the previous section (see Fig.~\ref{f43}). For our illustrative purpose we have plotted a numerical solution with $\da=0.04$, but it should be stressed that, for small enough $\da$, the curvature scale $H_0$ of the final de Sitter regime can be fixed at values arbitrarily smaller than the string scale ($H_0 \ll \ls^{-1}$), thus avoiding the need of taking into account $\ap$ corrections for an analytical description of this regime\footnote{In any case, the value of $H_0$, and then of $\da$, cannot be too high ($\da \ll 1$), otherwise it is unlikely for the dilaton to be trapped in a final stabilised configuration.}.

\begin{figure}[h!]
\centering
\includegraphics[width=7cm]{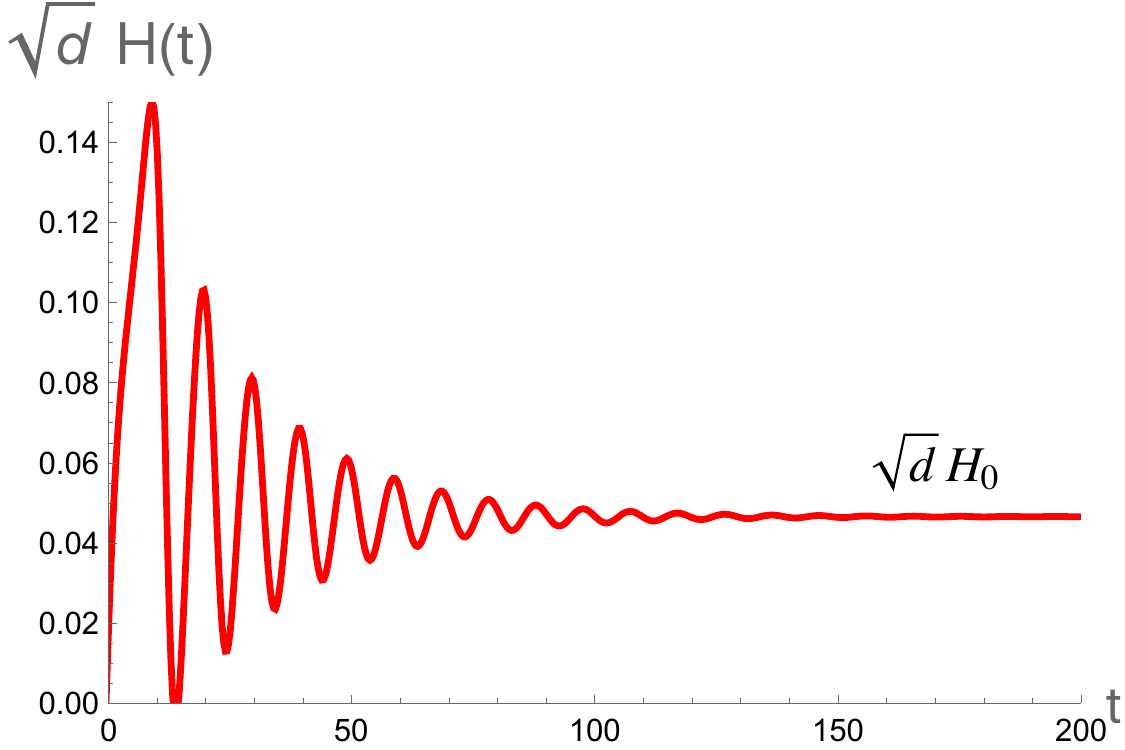}
~~~~~
\includegraphics[width=7cm]{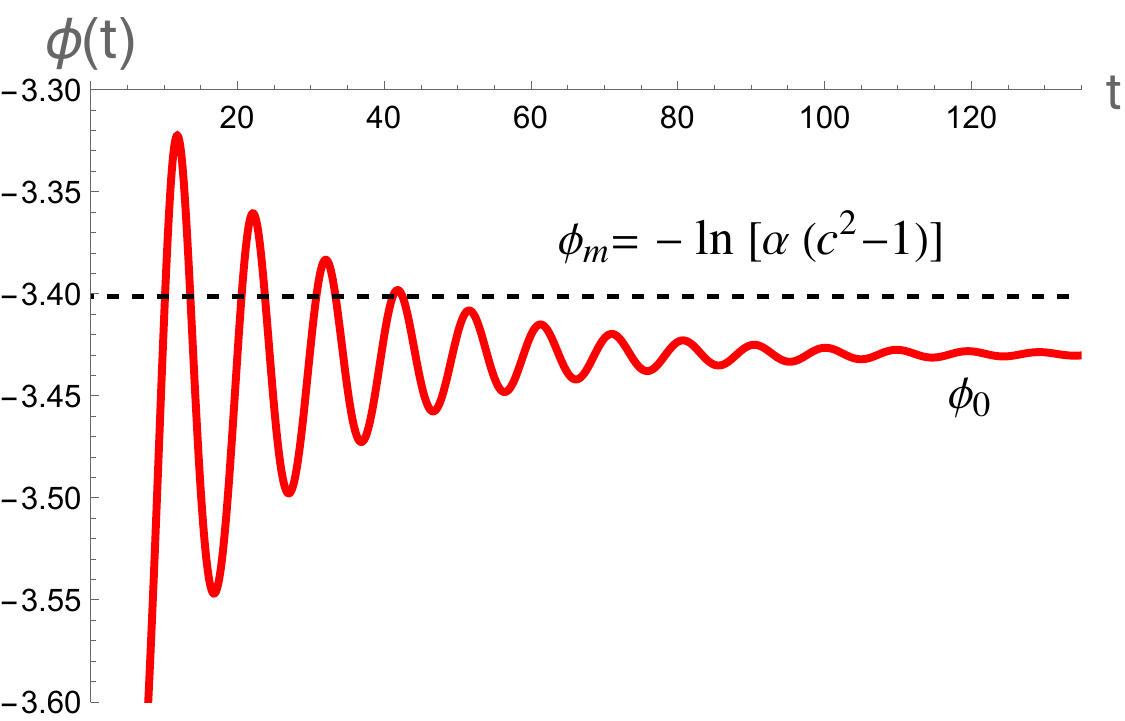}
\caption{Time evolution of $H(t)$ and $\phi(t)$ for the same numerical solution of Fig.~\ref{f44}. The final asymptotic regime is characterised by a constant value of $H_0>0$ and a constant dilaton $\phi_0 \not=\phi_m$.}
\label{f46}
\end{figure}

Another important difference from the case with $\da=0$ (not so evident, even if we compare the right plot of  Fig.~\ref{f44} with the similar one of Fig.~\ref{f43}), is that the final value of the stabilised dilaton, $\phi=\phi_0$, no longer coincides with the minimum of the potential, $\phi= \phi_m = - \ln [\a(c^2-1)]$. A better illustration of this effect can be obtained by plotting the time evolution of $H(t)$ and $\phi(t)$, shown in Fig.~\ref{f46}. The first plot shows that the final value of $H_0$ is non-vanishing, while the second plot clearly shows the difference between the stabilised value $\phi_0$ and the position of the local minimum $\phi_m$.

For an analytical discussion of this scenario in the final asymptotic regime and, in particular, for a physical interpretation of the difference between $\phi_0$ and $\phi_m$, let us now consider Eqs.~(\ref{17bis}) in the large time limit where we can neglect $\ap$ corrections. By evaluating such equations at the final attractor position, 
where $\phi=\phi_0$, $H=H_0$, $\dot \phi =0=\dot H$,
we than obtain the conditions
\beq
d \, H_0^2= -\left( \pa V\over \pa \phi \right)_{\phi_0}, ~~~~~~~~~~~~~~~
d \, (d-1) H_0^2= 2 \, V(\phi_0).
\label{410}
\eeq
They clearly show that $i)$ for a non-vanishing value of $H_0$ the final stabilised dilaton $\phi_0$ cannot be localised at the potential minimum $\phi_m$ (where $ \pa V/\pa \phi=0$); conversely, $ii)$ if the final attractor coincides with the minimum position, $\phi_0=\phi_m$, 
then we cannot avoid the final regime of standard evolution with decreasing curvature and a final value $H_0=0$ (consistently with the second equation (\ref{410}) and with the fact that, as discussed in Sect.~\ref{sec41}, in such a case the potential at the minimum is vanishing, $V_0 \equiv V(\phi_0)= V(\phi_m)=0$).

Combining the two equations (\ref{410}) we also obtain a condition on $\phi_0$ which can be written as
\beq
\left[ {\pa \over \pa \phi} \left(V e^{2 \, \phi\over d-1} \right) \right]_{\phi= \phi_0}=0,
\label{411}
\eeq
showing that the stabilised value $\phi_0$ corresponds to the minimum of an ``effective" potential $\ti V \sim V \exp\left[2 \phi/(d-1)\right]$. Interestingly enough, and consistently with the scenario of a stabilised dilaton, $\ti V(\phi)$ is nothing else that the dilaton potential for the action (\ref{12}) written in the E-frame (see e.g. \cite{9}). This is, as in the case of $V_0=0$, a consequence of the dilaton being a minimally coupled scalar only in the E-frame. We stress that the above effective potential contains an explicit dependence on the number of spatial dimensions, and this, as already noticed, may  affect the late-time post-bounce evolution of the gravi-dilaton background.

The difference between $\phi_0$ and $\phi_m$ can be analytically estimated (again, in the asymptotic regime of low enough curvature scales corresponding to $V_0 \ll 1$, $\da \ll 1$) by perturbatively expanding the potential around the minimum as
\beq
V(\phi)= V_0+{m^2\over 2}  \left (\phi- \phi_m\right)^2.
\label{412}
\eeq
By inserting this expression into the condition (\ref{411}), and solving for $\phi_0$, we obtain (to first order in $V_0/m^2$)
\beq
\phi_0 \simeq  \phi_m - {2 \, V_0\over m^2 (d-1)}+ {\cal O}\left(V_0^2\over m^4\right).
\label{413}
\eeq 
On the other hand we recall that for our potential (\ref{41}), and for small values of $\da$, we have  $V_0\simeq  A \da  (c^2-q) \exp(-c^2/\b)$. By computing $m^2=(\pa^2 V/\pa \phi^2)_{\phi=\phi_m}$, and  expanding the ratio $V_0/m^2$ in powers of $\da$ we obtain
\beq
{V_0\over m^2} = \da \, {c^2\over 2 \, (c^2-1)^2} +{\cal O}(\da^2),
\label{413a}
\eeq
so that for $\da \ll 1$ we have also $V_0/m^2\ll1$. In this approximation 
we can then obtain an analytical estimate of the de Sitter curvature scale $H_0$ in terms of $V_0$ and we find,  from Eqs.~(\ref{410}), (\ref{413}):
\beq
H_0 \simeq \left[ 2 \, V_0\over d \, (d-1)\right]^{1/2} \left[1+ {\cal O}\left(V_0\over m^2\right)\right].
\label{414}
\eeq

The range of initial conditions compatible with a final attractor corresponding to a stabilised dilaton and a final de Sitter geometry will be discussed in Sect. \ref{sec43}. The only difference from the previous case is that  $\da \not=0$ and $V_0 \not =0$.

\subsection{A numerical study of the initial conditions}
\label{sec43}

Here we provide a numerical study of the initial conditions that are necessary  for reaching a post-bounce phase with a stabilised dilaton. To this purpose we restrict ourselves to the approach presented at the end of Sect.~\ref{41} with the potential given in Eqs.~\eqref{41} and \eqref{41a} in which, we recall, we have set $\a'=1$. We also limit the discussion to the case $d=3$ although similar conclusions hold for generic $d$.

For the numerical code we set the initial conditions at $z_{in}=0.01$ and have checked that, as long as $z_{in} \ll 1$, the results are independent of $z_{in}$ provided we keep $\kappa$ fixed. We will then discuss how the allowed range of $\kappa$ compatible with a stabilised dilaton varies as a function of $A$, $\beta$ and $q$ while keeping fixed values for the other parameters i.e. $\a=10$, $c=2$ and $\delta=0.01$. Concerning this last choice, we anticipate that the  attraction basins do not appreciably change if $\delta=0$, so that the same discussion can be applied also to the case $V_0=0$, corresponding to the FLRW post-bouncing phase of Sect. \ref{sec41}.
On the other hand, if $\delta$ is so large that the local minimum has a height comparable to the peaks of $V$, the basin of attraction can be considerably reduced.

\begin{figure}[ht!]
\centering
\includegraphics[width=16cm]{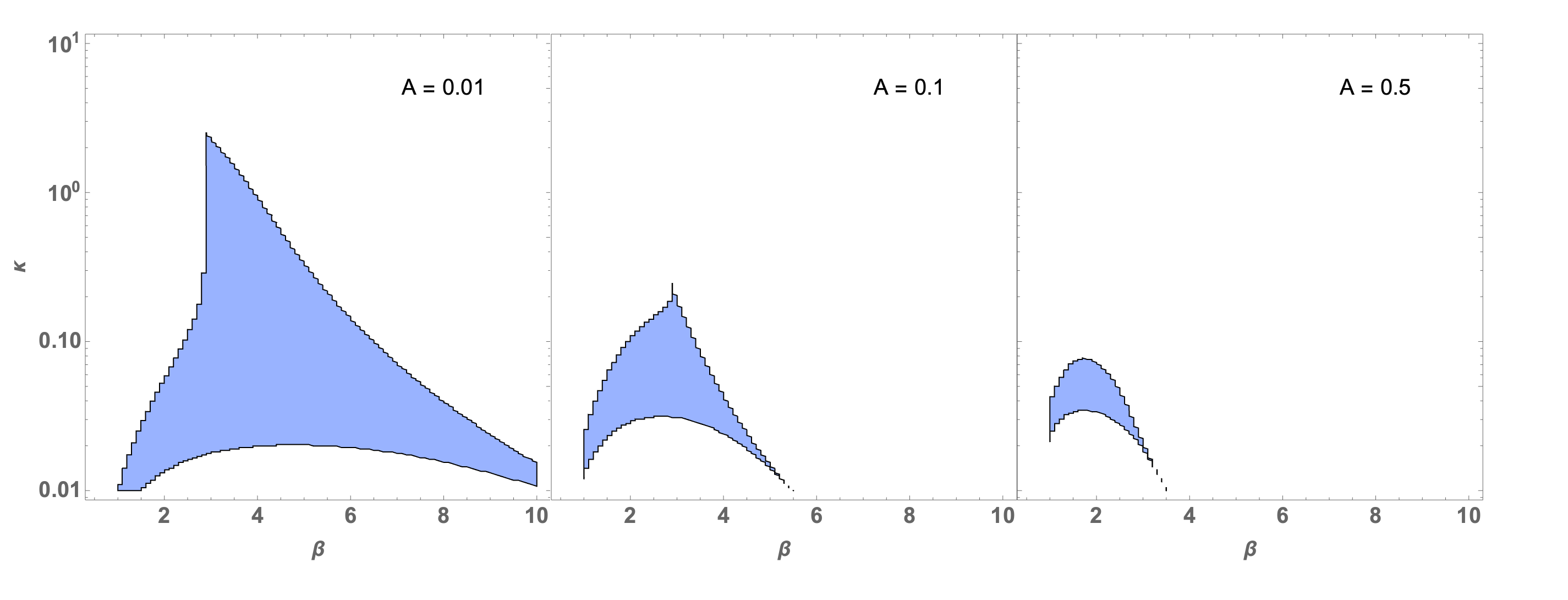}
\caption{The blue regions represent the basins of attraction for a post-bounce phase with stabilised dilaton, with the initial condition $z=0.01$ and $\kappa$ ranging as a function of $\beta$. We have adopted a dilaton potential \eqref{41} with $V_0>0$ and fixed the parameters of the potential  to be $\a=10$, $c=2$, $\delta=0.01$ and $q=1$. The magnitude of the potential, controlled by $A$, varies from smaller (left) to higher (right) values.}
\label{fig:basin1}
\end{figure}

Our first results are shown in Fig.~\ref{fig:basin1}, where we have considered the runaway potential with $q=1$ and varied $\beta$ and $A$. The range of $\kappa$ leading to a stabilised dilaton is illustrated by the blue regions reported in the figure, showing that the lower the magnitude of the potential, the larger is the basin of attraction.

In particular, for $A=0.01$, $\kappa$ may vary of about two orders of magnitude when $\beta=3$. This roughly corresponds to an initial amplitude of the dilaton in the range $\phi\in\left[-15,-9.5\right]$. This range drops dramatically to $\phi\in\left[-14.6,-13.6\right]$ when $A$ increases to $0.5$. This case also exhibits a shift of the peak value of $\beta$ to the lower value $\beta=2$.

\begin{figure}[ht!]
\centering
\includegraphics[width=16cm]{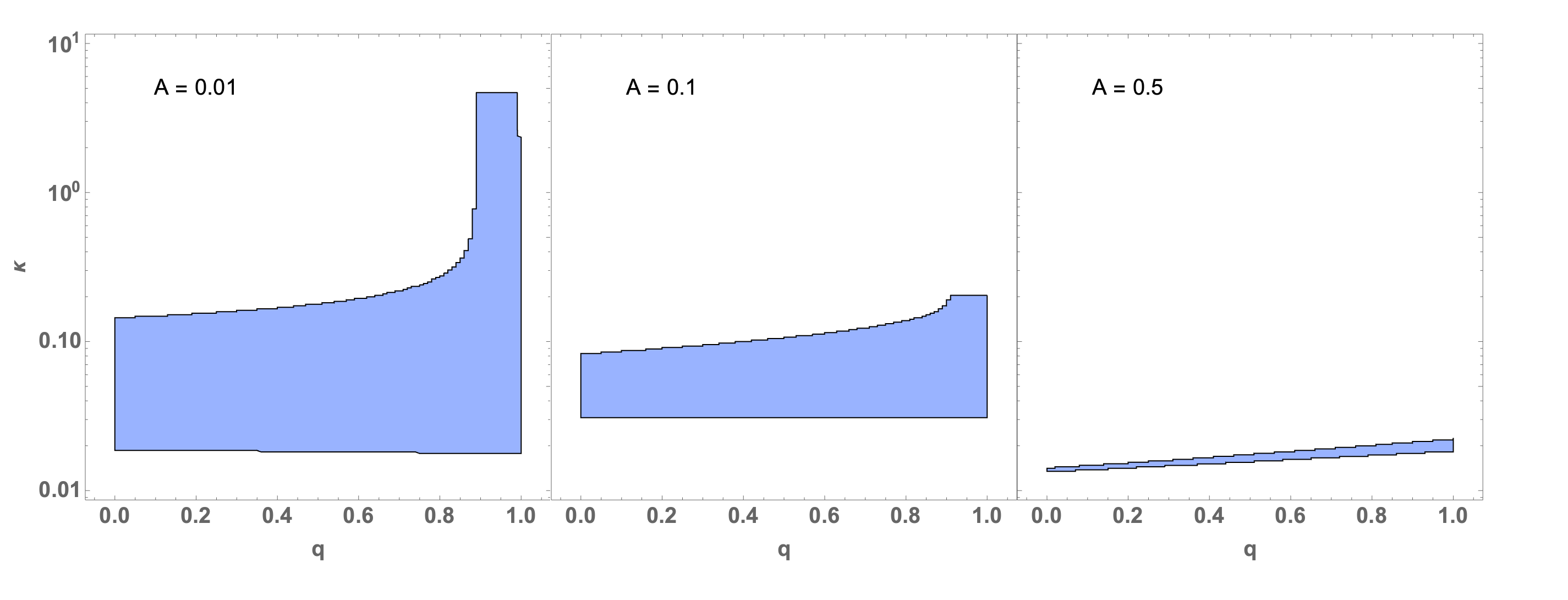}
\caption{The blue regions represent the basins of attraction of the post-bounce phase with a stabilised dilaton with  initial condition $z=0.01$ and $\kappa$ ranging as a function of $q$. We have adopted the dilaton potential \eqref{41} with $V_0>0$ and fixed the parameters of the potential  to be $\a=10$, $c=2$, $\delta=0.01$ and $\beta=3$. The magnitude of the potential controlled by $A$ varies from smaller (left) to higher (right) values.}
\label{fig:basin2}
\end{figure}

To give an alternative discussion, in Fig.~\ref{fig:basin2} we have set $\beta=3$ and varied $\kappa$ as a function of $A$ and $q$, exploring then the basin of attraction for different behaviours of the potential in the limit $\phi \ra +\infty$.
Independently of the value of $A$, the general trend is that a larger range of $\kappa$ is achieved when the runaway potential tends to vanish at $\phi=+\infty$. On the contrary, the range of $\kappa$ is suppressed when the potential in the same limit reaches its maximum.

Let us finally note that, for $d=3$, if the initial  value of $\kappa$  is outside the allowed range of Fig. \ref{fig:basin2} then the dilaton cannot be stabilised, it ``bounces back" to $-\infty$, and the final background evolution asymptotically corresponds to the time reversed of the initial one (i.e. to the second one of the three cases mentioned in Sect. \ref{sec4}, and illustrated in more detail in  Appendix \ref{appA}). This happens because for $d=3$ the (naively computed) effective potential in the E-frame never goes to zero as $\phi\rightarrow +\infty$. However, for higher values of $d$, or for different choices of the potential, the situation can be different and there are two possibilities for a non-stabilized dilaton: the  dilaton runaway scenario (with growing dilaton both before and after the bounce) corresponding to the first case of Sect.~\ref{sec4}, and the previous case  of a bouncing dilaton.

\begin{figure}[ht!]
\centering
\includegraphics[width=16cm]{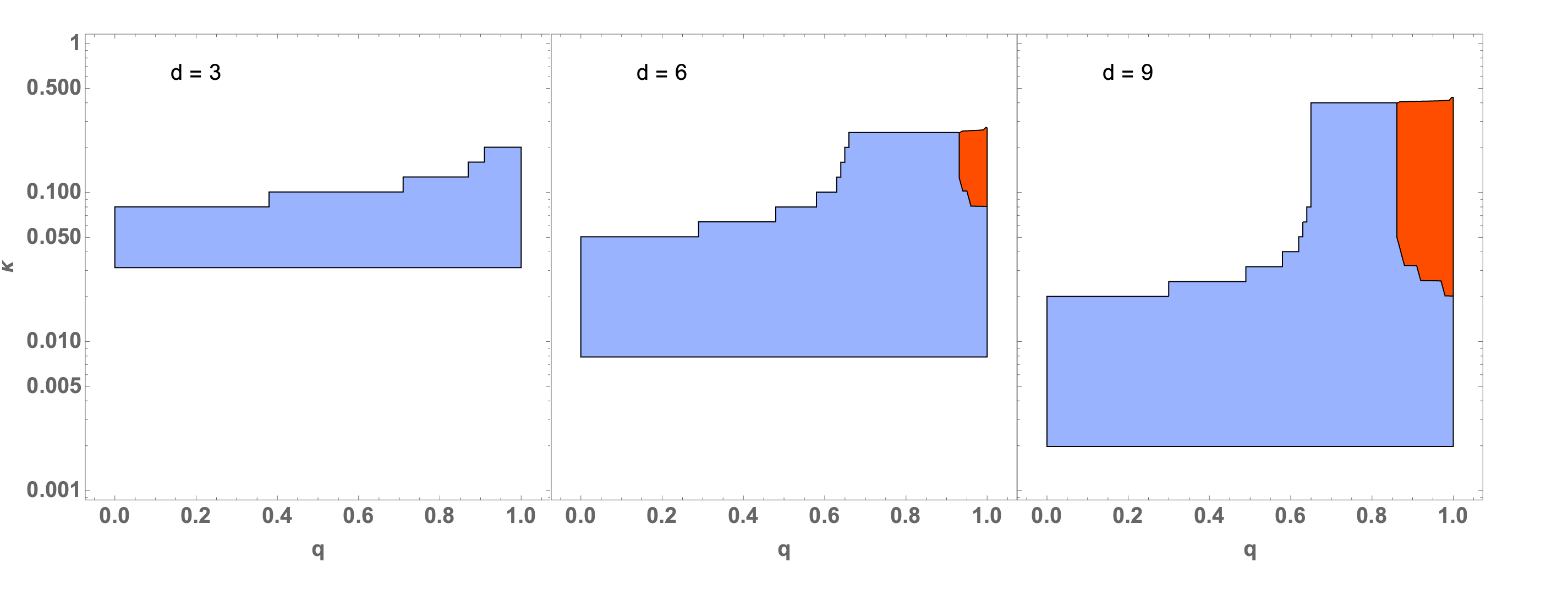}
\caption{The same as Fig.~\ref{fig:basin2}, but we have fixed $A=0.1$ and varied the number of spatial dimensions. The blue regions represent the basins of attraction of the post-bounce phase with a stabilised dilaton. The red regions correspond to initial conditions leading to a post bounce phase  with unbounded growth of the dilaton. In the white regions, the final post bounce phase corresponds to the time reversal of the initial one.}
\label{fig:higherdimension}
\end{figure}

This is illustrated in Fig.~\ref{fig:higherdimension}, where we have varied the number of spatial dimensions keeping fixed $A=0.1$ and with all  other parameters fixed at the same values as in Fig.~\ref{fig:basin2}. The red regions, appearing for $d>3$ and lying outside the attraction basin, correspond to initial conditions leading to a final unbounded growth of the dilaton.

\section{Dilaton stabilisation and isotropisation by $\alpha'$ corrections and $V\ne 0$}
\label{secaniso}

In this Section we consider the case of a generically anisotropic initial (pre-bouncing) cosmology described by a Bianchi-I type metric, and address the question of whether, in the presence of a potential, it can also admit late time attractors with a stabilised dilaton. Furthermore, we can ask whether such attractors, when they exist, are isotropic. This is, of course, an important question since we do not want to fine-tune the initial conditions to be isotropic: like in the standard inflationary scenario we would like initial anisotropy to be washed out at late times. We remind the reader that the low-curvature Kasner-like solutions of string cosmology do not have such a property: the ratio between the initial shear and the volume-expansion rate remains essentially constant in the lowest-order solutions \cite{Kunze:1999xp}.

Taking into account $\alpha'$ corrections and a dilaton potential the situation can change. We know that, in principle, there are multi-trace terms in the original HZ action (and consequently in our ``Hamiltonian" $h(z_i)$). Such terms, by coupling different $H_i$'s can mimic effective ``viscosity terms", which are known to be able to wash out (or limit the growth of) the initial anisotropies (see e.g. \cite{Quintin} and references therein). However, we have not made yet a systematic study of this possibility, also because such multi-trace terms are supposed to emerge only at order $\alpha'^3$ (or higher) in the curvature expansion \cite{HZ}. Fortunately, late-time isotropisation of an initial anisotropic metric can also occur as a result of the presence of $V(\phi)$ alone, which induces a non-linear dependence upon $\sum \beta_i$.

It is known, in fact, that the presence of homogeneous and isotropic sources (like, for instance, the effective radiation fluid describing the back-reaction of the amplified perturbations) may lead to a final asymptotic regime of isotropic expansion even starting from an initial regime with both expanding and contracting dimensions (see e.g. \cite{24}). In our context the role of the homogeneous source can be  played by the potential energy of the stabilised dilaton, and 
the implementation of such an isotropisation mechanism only requires starting from initial conditions compatible with the final stabilising attractor (see the discussion at the end of this section). 
In particular, the following simple result can be obtained from the relevant equations (\ref{16}):
\\
\\
{\bf Whenever there is a late-time attractor with constant $\phi$ and $z_i$ the attractor must be isotropic, i.e $z_i = z=z_0$, and consequently $H_i = H=H_0$.}
\\
\\
The proof is simple: If $\phi=\phi_0=$ const then $\fbp = - \sum_i H_i$ and the last two Eqs.~(\ref{16}) become, in the attractor limit,
\beq
 \left(\sum_j H_j\right) z_i  +{\pa V\over \pa \phi} = 0~~;~~~~~~~~~~~~~~~ \sum_i z_i H_i +  {\pa V\over \pa \phi} = 0 \, .
\label{FPan}
\eeq
The first set of equations tells us already that $z_i$, at the attractor, is independent of $i$, which is consistent with the last equation and gives
\beq
d \, z_0 \, H_0 = - \left(\pa V\over \pa \phi\right)_{\phi=\phi_0}.
\label{isoattr1}
\eeq 
Then the first of Eqs.~(\ref{16}) gives the further constraint:
\beq
d^2 H_0^2= 2 \left(h\right)_{z_i=z_0}+2 \, V(\phi_0).
\label{isoattr2}
\eeq 
The system of equations (\ref{isoattr1}),  (\ref{isoattr2}) (equivalent  to the previous Eqs.~(\ref{410}) if the attraction point is located in the low-energy regime) is not over-constrained, and typically admits a finite number of solutions for $\phi_0$ and $z_0$, and thus for $H_0$, once $h(z_i)$ is given.

The existence of such attractors is confirmed by a numerical study of Eqs.~(\ref{16}).
Under a suitable choice of the potential and of the initial conditions one can find indeed regular solutions described, in the parametric plane $\{\fbp, H_i\}$, by smooth curves (like those of Figures~\ref{f41},~\ref{f44}), connecting the initial anisotropic configuration to the same final fixed point attractor. Let us give immediately an explicit example 
for an anisotropic, Bianchi-I type gravi-dilaton background whose dynamical evolution is controlled by this simple, $\ap$-corrected, effective Hamiltonian:
\beq
h(z_i)= {1\over 2} \sum_i z_i^2 - c_2 {\ap\over 8}\sum_i z_i^4.
\label{51}
\eeq
For $c_2=2$, and in the isotropic $d$-dimensional limit, we recover the Hamiltonian (\ref{43}) used in the numerical examples of the previous sections. We also note that (after taking into account the overall factor $-1/2$ of difference in the definition of $h$) this model of Hamiltonian exactly coincides with the model of anisotropic $\ap$ corrections presented in Ref.\cite{Gasperini:2023tus} but {\it without} the multi-trace term considered in that paper.

For a better graphical illustration of the numerical solution it is convenient to consider a potential with a non-vanishing local minimum $V_0>0$, so that the final isotropic attraction point in the  $\{\fbp, H\}$
plane is different from the origin. We will thus consider a potential of the ``runaway" type with $\da =0.01$, and with all the the same values of the other parameters as those used in the plot of Fig.~\ref{f44} (except for the numerical values  $A=0.1$ and $\b=3$). Also, we shall work with a $d+n$-dimensional space, where $d$ dimensions are expanding with scale factor $a_1$ and $n$ dimensions are expanding with scale factor $a_2$, imposing, as initial conditions at $t\ra -\infty$, $H_i \ra 0$, that the solution satisfies 
the low-energy pre-big bang equation $dH_1^2+ nH_2^2 = \fbp^2$, with $H_1 \not= H_2$ and $H_1>0$ , $H_2>0$.
In this limit, the solution can be explicitly written as \cite{9}
\beq 
a_i \sim (-t)^{-\gamma_i}, ~~~~~\gamma_i>0,  ~~~~~ \sum \gamma_i^2=1, ~~~~~ H_i ={\gamma_i\over t}, ~~~~~ \phi \sim - \left(\sum \gamma_i +1\right) \ln (-t),
\label{51a}
\eeq
and we can define a constant $\kappa$, as in Sect.~\ref{sec41}, which controls the initial conditions and is determined by the initial values of $\phi$, $H_1$ and $H_2$ as follows:
\beq
\kappa = e^\phi \left[d \, H_1 +n \, H_2\over \sqrt{d+n}\right]^{-(1 + d \, \gamma_1+n \, \gamma_2)}\,\sim \,e^\phi \left[d \, z_1 +n \, z_2\over \sqrt{d+n}\right]^{-(1 + d \, \gamma_1+n \, \gamma_2)}\,,
\label{51b}
\eeq
where the last equality holds in the limit $t\rightarrow -\infty$, in analogy with what discussed in the isotropic case.
The initial anisotropy is imposed by choosing $\gamma_1= \sqrt{(1 - n  \, \epsilon) /(d+n)}$ and $\gamma_2=\sqrt{ (1 + d \,   \epsilon)/(d+n)}$, satisfying, for any $\epsilon$, the Kasner condition of (\ref{51a}) i.e. $d \gamma_1^2 + n\gamma_2^2 = 1$.

\begin{figure}[t]
\centering
\includegraphics[width=4 cm]{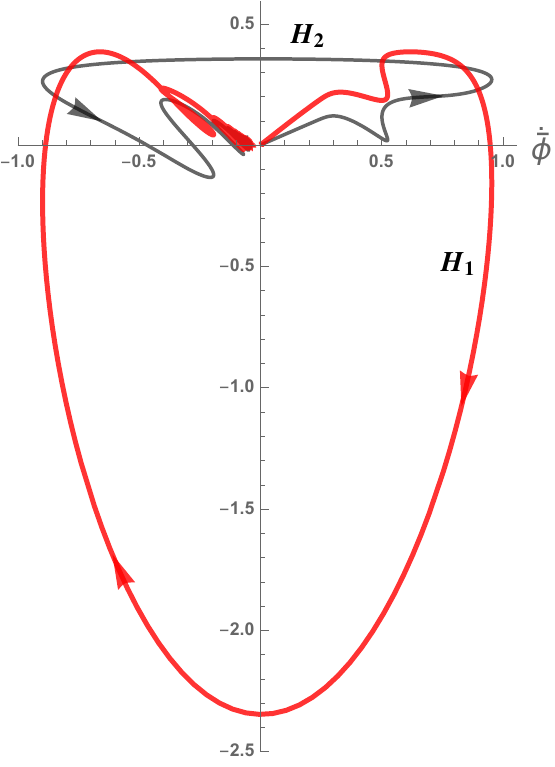}~~~~~~~~
\includegraphics[width=8.5cm]{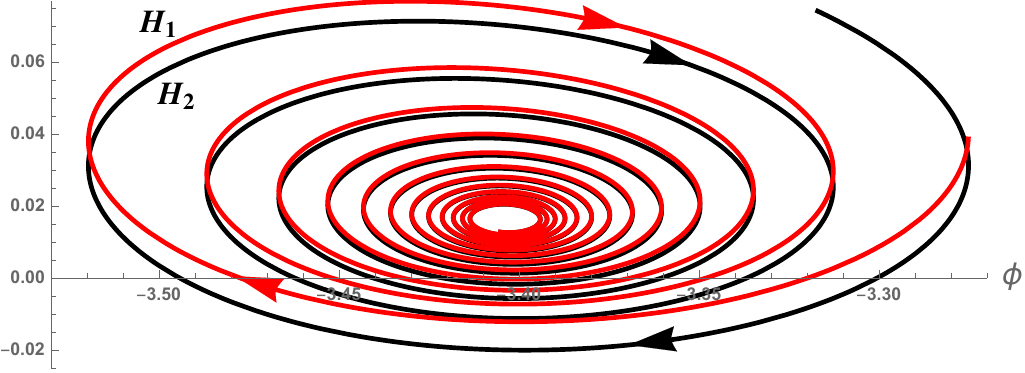}
\caption{Left: parametric plot of a numerical solution of Eqs.~(\ref{16}) with the potential (\ref{41}) and the Hamiltonian (\ref{51}), written for  $c_2=2$, $\ap=1$. We have assumed an anisotropic configuration with $d$ coordinates with momentum $z_1$ and $n$ coordinates with momentum $z_2$. The red curve describes the evolution of the $d$ Hubble parameters $H_1$, the black curve of the $n$ Hubble parameters $H_2$. The initial anisotropy is specified by $\epsilon = 0.3$ and we have chosen an initial parameter (see Eq.~(\ref{51b})) $\kappa=0.073$. Finally, we have set $d=1$, $n=2$, and used the following numerical values for the potential:
 $A=0.1$, $\a=10$, $\b=3$, $c=2$, $\da =0.01$, $q=1$. Right: parametric plot of $H_1(\phi)$ (red curve) and $H_2(\phi)$ (black curve) for the same numerical solution, in the asymptotic range  of large positive times. The attraction point is localized at $H_0 \simeq 0.016$, $\phi_0\simeq -3.4$, according to Eqs.~(\ref{413})--(\ref{414}).}
\label{f51}
\end{figure}

 The numerical results are illustrated in Fig.~\ref{f51} for $d=1$ and $n=2$ (however, the results is qualitatively the same for different choices of $d$ and $n$, for instance $d=3$ and $n=6$). As clearly shown in the figure, in spite of the very different (largely anisotropic) behaviour of the $d$ and $n$ spatial dimensions both before and during the bounce, after the bounce they all converge (after an oscillating epoch) to the same attraction point, controlled by the dilaton trapped in the potential minimum. 
Note that the two parametric curves for $H_1$ and $H_2$ plotted in the left sector of Fig.~\ref{f51} are not topologically equivalent in the sense explained at the end of Sect.~\ref{sec3}, as one is turning clockwise, the other anti-clockwise, in the parametric plane $\{\fbp, H\}$. This difference from the solutions of Sect.~\ref{sec3} is due to the presence of the dilaton potential.

 The final oscillations, as well as the isotropic  convergence towards the same final attractor $\{\phi_0, H_0\}$, are clearly displayed if we restrict the parametric plots to the large time limit, obtaining the behaviour of $H_1(\phi)$ and $H_2(\phi)$ illustrated in the right sector of Fig.~\ref{f51}. Given the numerical values of the potential parameters, and using Eqs.~(\ref{413})--(\ref{414}), we can easily compute the coordinates $\phi_0$, $H_0$ of the attraction point, and check that they coincide with the result shown in the figure.

\begin{figure}[ht!]
\centering
\includegraphics[width=15cm]{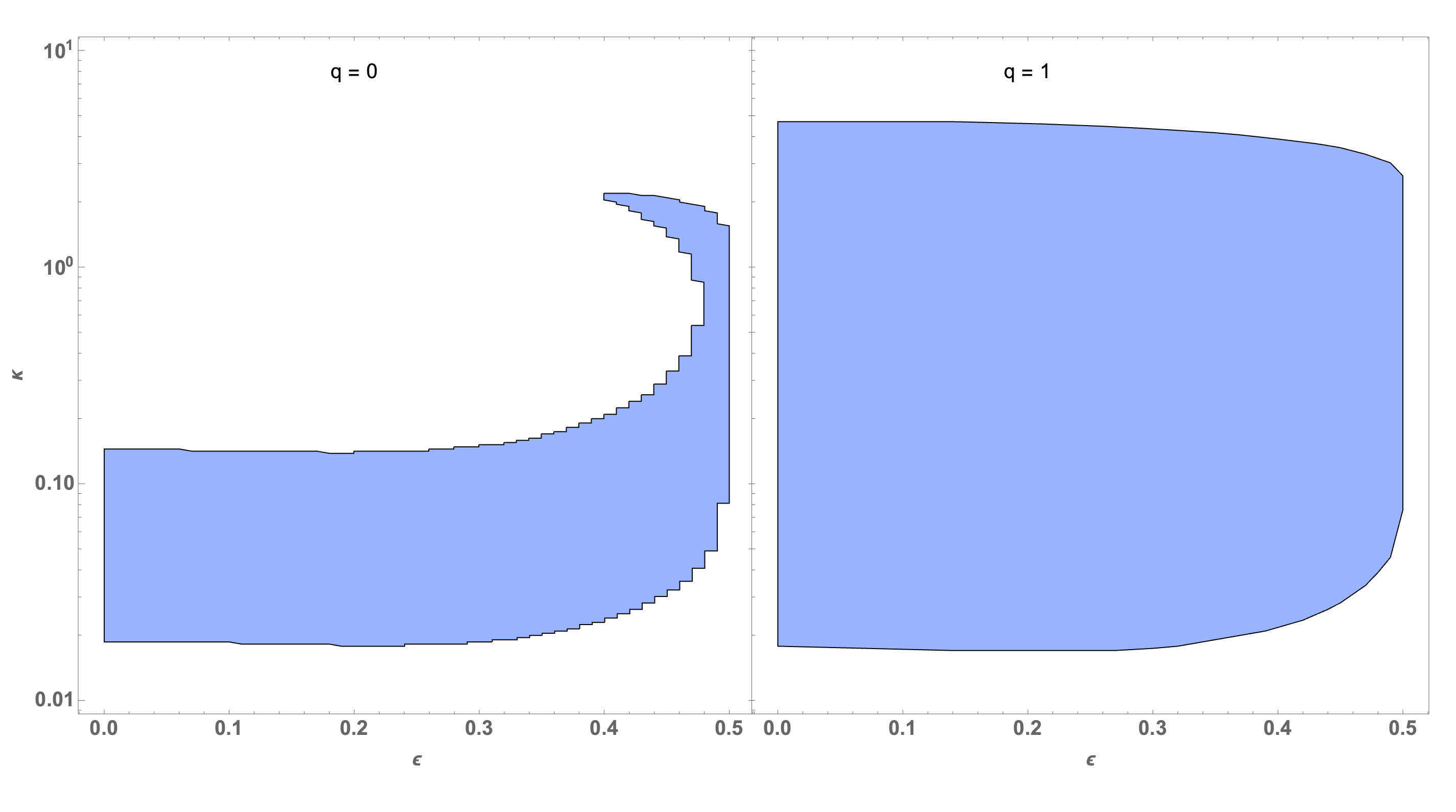}
\caption{The blue regions represent the basins of attraction of the post-bounce phase with a stabilised dilaton and spatial isotropisation in the parametric plane $\left\{\epsilon,\kappa\right\}$ in $1+2$ dimensions. Here we have adopted the dilaton potential \eqref{41} with $V_0>0$ and fixed the parameters of the potential  to be $A=0.01$, $\a=10$, $c=2$, $\delta=0.01$ and $\beta=3$. We show the separate results  for $q=0$ (left) and $q=1$ (right).}
\label{fig:basin3}
\end{figure}

Let us now discuss the basin of attraction for this class of anisotropic backgrounds following the same discussion presented in Sect.~\ref{sec43}. To this purpose we will use of the definition of $\kappa$ of Eq. \eqref{51b} to study  the initial conditions in the parametric plane $\left\{ \epsilon,\kappa \right\}$. For what concerns the potential, following the results obtained in the isotropic case, we will consider the illustrative example with $A=0.01$, $\a=10$, $\beta=3$, $c=2$ and $\delta=0.01$, choosing $d=1$ and $n=2$ for the geometrical shear. This limit $\epsilon$ to range from $0$ (isotropic limit) to $0.5$ (highest anisotropy) in order for the $\gamma_i$'s to be real. The results we obtain are shown in Fig.~\ref{fig:basin3} for the cases $q=0$ and $q=1$.

In both cases the basin of attraction remains quite stable, regardless of the amplitude of the initial geometrical shear. The case of a potential reaching its maximum at $\phi=+\infty$ ($q=0$) looks quite intriguing, since in this case a shift towards slightly higher and multi-valued regions of $\kappa$ emerges. However, the overall permitted amplitude of  $\kappa$ is quite insensitive to the degree of initial anisotropy: hence, the mechanism proposed to get a suitable post-bounce phase with a stabilised dilation looks pretty robust  to guarantee also a final isotropisation, regardless of the amplitude of the anisotropy in the pre-bounce perturbative vacuum.


\section{Summary and outlook}
\label{sec6}

Let us start by summarizing the main results of this paper. 
We started by making more precise and general the reformulation given in \cite{Gasperini:2023tus} of the Hohm-Zwiebach approach to ``classical"  (i.e at lowest order in the string loop expansion but at all orders in $\ap$) string cosmology. We did this by appealing to the Routhian formulation of dynamical systems in which part of the degrees of freedom are treated in a Hamiltonian approach (through the standard Legendre transform) while the rest of the variables is kept at the Lagrangian level. This approach is known to be most useful when some of the former degrees of freedom are cyclic variables leading to corresponding conservation laws. This is precisely the situation for the $d$ scale-factors of a Bianchi I cosmology to which the HZ treatments applies. We have also stressed that, in this case, when the Hamiltonian of the system obeys the conditions spelled out in \cite{Gasperini:2023tus}, the regular bouncing  solutions preserve a subgroup of the $SFD \otimes T$ global symmetry, unlike the lowest-order singular ones that break it spontaneously. These regular solutions can be classified as being ``self-dual".

We have then added to the Lagrangian a non-perturbative, local dilaton potential. In this case, having assumed the principle of ``Asymptotic Past Triviality" formulated in \cite{Buonanno:1998bi}, the above-mentioned conservation laws are still valid in the asymptotic past but are explicitly broken as soon as the potential starts to be felt. As a result we are still able  to give  the initial data in terms of the asymptotically conserved quantities.

After having specified the potential in terms of a few parameters we have studied numerically, and in certain simple situations analytically, the evolution of the system at late times. The main results of this analysis can be summarized as follows:
\begin{itemize}
    \item The presence of the potential (always taken not exceeding $V$ of order one in string units) does not seem to affect the existence or non-existence of regular bounces: this seems to depend essentially only on the form of the kinetic part of the Hamiltonian as discussed in \cite{Gasperini:2023tus}. Nonetheless, the potential can affect the cosmological evolution both during and, even more dramatically, after the bounce.
    \item For choices of the Hamiltonian leading to a regular bounce, the late-time cosmology is determined by the shape of the potential and by the initial conditions. Several cases are possible, not all equally interesting for physics.
    \item In the isotropic case we have seen three possible behaviors at late time: $i)$ the dilaton can go through the potential barriers and ends up again in the same dual asymptotic solution as the one encoutered in the $V=0$ case; $ii)$ it can be ``reflected" by the potential and go back to $- \infty$ (see Appendix~\ref{appA} for details), in which case the final cosmology is just the time-reversal of the initial one; $iii)$ finally, and most interesting,  the dilaton can be stabilized at the minimum of an effective potential $\ti V$ 
    (not necessarily identical to the given $V$) and the late time evolution is either a matter-dominated FLRW decelerating cosmology, or a de-Sitter like accelerated expansion, depending on the value of the potential at the minimum. 
    \item For anisotropic initial conditions the situation is similar with an extra welcome feature: whenever the dilaton is stabilized at late times, the attractor is isotropic so that the regular bounce, with the help of the potential, washes out any initial anisotropy. 
\end{itemize}

 For the future, several interesting questions remain to be addressed:
 \begin{itemize}
     \item To implement in these scenarios a stabilizing mechanism for the internal compact dimensions (assumed to satisfy the usual $T$-duality symmetries).
     \item To add the Kalb-Ramond $B_{\mu\nu}$ field (possibly in its axionic dual form) in order to exploit the full $O(d,d)$ symmetry of the HZ action. We recall that the axion plays an important role in the traditional pre-big bang scenario via the curvaton mechanism \cite{Bozza1,Bozza2}. It is not clear whether, in this new context, such a mechanism is still needed (see below).
     \item 
     Computing perturbations in these new scenarios is an important but non trivial problem since it requires to break the abelian isometries of the homogeneous solutions.  On the other hand,  bouncing scenarios, unlike ordinary inflation, do not suffer from a transplanckian problem (see e.g. \cite{Brandenberger:2012aj}) so that initial conditions can be given in the perturbative, asymptotically trivial, regime. In particular, if the final attractor is de-Sitter-like (as in Sect. (\ref{sec42})), we may expect various branches in the spectrum of abiabatic scalar and tensor perturbation (neglecting the isocurvature axionic perturbations discussed in \cite{Bozza1,Bozza2}): perturbations going out of the horizon during the initial perturbative phase of dilaton-driven inflation (leading to a steep blue spectrum \cite{8}), those coming from the (relatively short) high curvature bounce and, finally, perturbations generated by the final frozen dilaton de Sitter phase and expected to be nearly scale-invariant.
 \end{itemize}

At a more conceptual level we wish to emphasize again the two most important open issues: 

1. Can the Lagrangians, Hamiltonians, or Routhians that implement a regular bounce correspond to the dimensional reduction of some general covariant action to the case of an homogeneous background with $d$ abelian isometries? Constructing such an action, at least perturbatively in the spatial gradients, would be also most important phenomenologically, in particular for the study of perturbations.

2. This whole approach is deeply routed in the duality symmetries of string theory but we are still lacking tools for finding out whether the particular duality-invariant models that looks phenomenologically promising do follow from any specific consistent string theory.


\appendix
\section{Appendix. From expanding pre-big bang to contracting post-big bang}
\label{appA}

As discussed in Sect.~\ref{sec4}, one of the possible effects of the potential (\ref{41}) is that of producing an effective potential barrier which stops the growth of the dilaton and ``reflects back" the dilaton towards the large negative values of the perturbative string vacuum. 

Such an effect, combined with the curvature regularisation due to the $\ap$ corrections, leads to a new interesting type of bouncing scenario where the final background is no longer described by the duality-transformed initial solution (indeed, duality is broken by the presence of the potential), but it nevertheless corresponds to the time-reversed version of the initial solution. Hence, if we start with an initial (isotropic, $(d+1)$-dimensional) pre-big bang configuration describing growing dilaton, growing curvature, and accelerated expansion ($\dot a >0$, $\ddot a >0$):
\beq
a \sim (-t)^{-1/\sqrt{d}}, ~~~~~ H = {1\over \sqrt{d} \, (-t)} >0, ~~~~~
\phi \sim -\left(1 + \sqrt{d}\right) \ln (-t), ~~~~~
\fbp = \sqrt{d} \, H>0, ~~~~~
t \ra -\infty,
\label{a1}
\eeq
(see also Eq.~(\ref{earlyt})), we end up with a post-bounce configuration asymptotically describing decreasing dilaton, decreasing curvature, and decelerated contraction ($\dot a<0$, $\ddot a >0$):
\beq
a \sim (t)^{-1/\sqrt{d}}, ~~~~~ H = -{1\over \sqrt{d} \,(t)} <0, ~~~~~
\phi \sim -\left(1 + \sqrt{d}\right) \ln (t), ~~~~~
\fbp = \sqrt{d} \,  H<0, ~~~~~
t \ra +\infty.
\label{a2}
\eeq

\begin{figure}[t]
\centering
\includegraphics[width=4.5 cm]{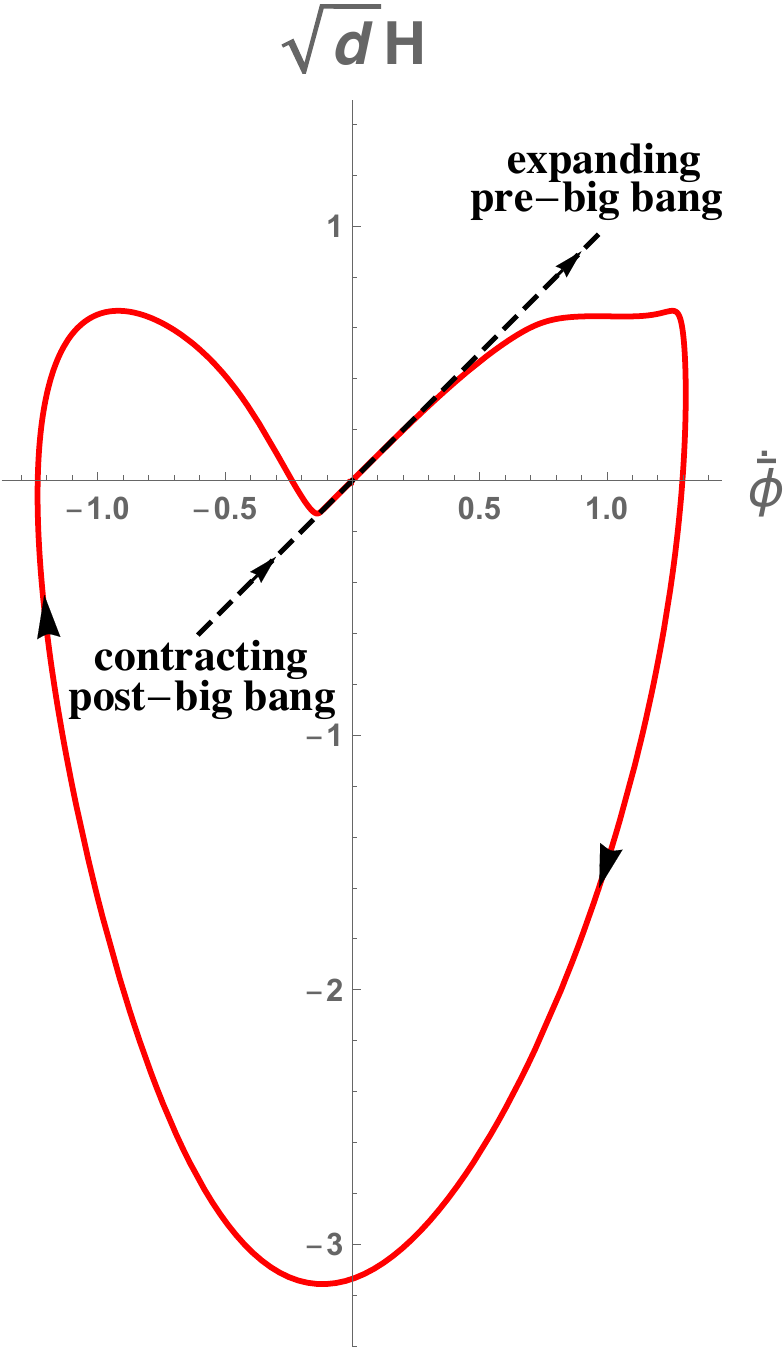}~~~~~~~~
\includegraphics[width=6 cm]{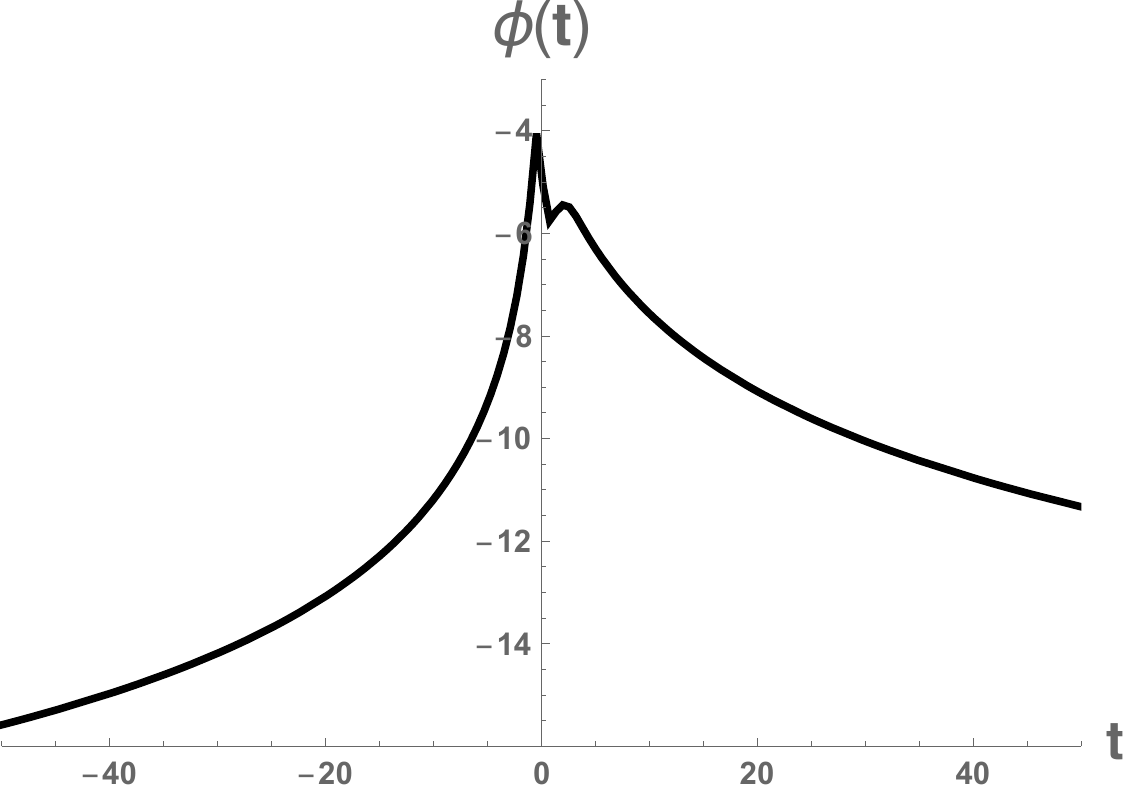}
\caption{Left: parametric plot obtained with a numerical solution of Eq.~(\ref{17bis}) with the potential (\ref{41}) and the Hamiltonian (\ref{43}). We have set $d=3$, $\ap=1$ for the Hamiltonian, $A=1/3$, $\a=10$, $\b=3$, $c=2$, $\da=0.01$, $q=1$ for the potential, and $\kappa=0.0075$ for the initial conditions at $t=-200$. The black dashed bisecting line describes the initial evolution  (\ref{a1}) expanding from the perturbative regime (upper-right quadrant), and the final time-reversed contracting evolution  (\ref{a2}) (lower-left quadrant). Right: time evolution of the dilaton according to a numerical integration of (\ref{17bis}) performed with the same parameters and the same initial conditions.}
\label{fa1}
\end{figure}

A simple illustration of this scenario can be provided by a numerical integration of Eqs.~(\ref{17bis}) with appropriate values of the initial conditions, which in our case are controlled by the parameter $\kappa$ of Eq.~(\ref{49bis}), defined at large negative times by $\kappa=e^\phi \left(\sqrt{d} \, H\right)^{-(1+\sqrt{d})}$.
According to Eq.~(\ref{a1}) this parameter goes to a constant for $t \ra -\infty$. Such a constant depends on the initial value of the dilaton and, as discussed in Sect.~\ref{sec41}, it is equivalent (but more convenient) to specify initially the value $\kappa$ rather than that of $\phi$. For initial values of $\kappa$ small enough with respect to the parameters controlling the height of the potential, the numerical integration gives the results reported in Fig.~\ref{fa1} (see the caption for the used numerical values).

In the left panel we have the parametric plot describing a smooth, regular transition from the initial configuration (\ref{a1}) to the final configuration (\ref{a2}). The corresponding evolution of the dilaton, which reaches a maximum and then bounces back to the asymptotically perturbative regime, is shown in the right panel of the figure. Note that the red curve describing the parametric evolution of $H(\fbp)$ starts and ends at the origin, is still turning clockwise in the parametric plane, but it is not topologically equivalent to the curves given in the plots of Sect.~\ref{sec3} and Sect.~\ref{sec41}, (always associated with a final expanding regime). Indeed, in the parametric plot of of Fig.~\ref{fa1} the vector connecting the origin to a point on the red curve undergoes a clockwise rotation
of $\pi$ (and not $3\pi/2$) as one goes from the beginning to the end of the curve.

\section*{Acknowledgements} 
GF acknowledges support by the FCT under the program {\it ``Stimulus"} with the grant no. CEECIND/ 
04399/2017/CP1387/CT0026 and through the research project with ref. number PTDC/FIS-AST/ 0054/2021. GF is also member of the Gruppo Nazionale per la Fisica Matematica (GNFM) of the Istituto Nazionale di Alta Matematica (INdAM).
PC, MG, EP and LT are supported in part by INFN under the program TAsP: {\it ``Theoretical Astroparticle Physics"}. MG and LT are also supported by the research grant number 2017W4HA7S {\it ``NAT-NET: Neutrino and Astroparticle Theory Network"}, under the program PRIN 2017 funded by the Italian Ministero dell'Universit\`a e della Ricerca (MUR). PC, GF, MG, EP and LT wish to thank the hospitality of CERN, Theory Department, where part of this work has been carried out. GV would like to acknowledge useful discussions with J. Maharana. Finally, we would also like to thank R. Brandenberger, R. Brustein, J. Quintin and B. Zwiebach for useful comments on a preliminary version of this paper.



\begin{thebibliography}{999}
\newcommand{\bb}{\bibitem}


\bibitem{Meissner:1991zj}
K.~A.~Meissner and G.~Veneziano,
Symmetries of cosmological superstring vacua,
Phys. Lett. B \textbf{267} (1991) 33.

\bibitem{Meissner:1991ge}
K.~A.~Meissner and G.~Veneziano,
Manifestly O(d,d) invariant approach to space-time dependent string vacua,
Mod. Phys. Lett. A \textbf{6} (1991) 3397
[arXiv:hep-th/9110004 [hep-th]].

\bb{1b}M. Gasperini and G. Veneziano, 
O(d, d) covariant string cosmology,
Phys. Lett. B {\bf 277}
(1992) 256 [hep-th/9112044].

\bibitem{Gasperini:1993hu}
M.~Gasperini and G.~Veneziano,
Inflation, deflation, and frame independence in string cosmology,
Mod. Phys. Lett. A \textbf{8} (1993), 3701-3714
doi:10.1142/S0217732393003433
[arXiv:hep-th/9309023 [hep-th]].

\bibitem{Maharana:1992my}
J.~Maharana and J.~H.~Schwarz,
Noncompact symmetries in string theory,
Nucl. Phys. B \textbf{390} (1993) 3-32
[arXiv:hep-th/9207016 [hep-th]].

\bibitem{Sen:1991zi}
A.~Sen,
O(d) x O(d) symmetry of the space of cosmological solutions in string theory, scale factor duality and two-dimensional black holes,
Phys. Lett. B \textbf{271} (1991) 295-300.

\bibitem{Meissner:1996sa}
K.~A.~Meissner,
Symmetries of higher order string gravity actions,
Phys. Lett. B \textbf{392} (1997) 298-304
[arXiv:hep-th/9610131 [hep-th]].

\bibitem{Codina1}
T.~Codina, O.~Hohm and D.~Marques,
String Dualities at Order $\alpha'^{\,3}$,
Phys. Rev. Lett. \textbf{126} (2021) no.17, 171602
[arXiv:2012.15677 [hep-th]].

\bibitem{Codina2}
T.~Codina, O.~Hohm and D.~Marques,
General string cosmologies at order $\alpha'^{\,3}$,
Phys. Rev. D \textbf{104} (2021) no.10, 106007
[arXiv:2107.00053 [hep-th]].

\bibitem{Veneziano:1991ek}
G.~Veneziano, Scale factor duality for classical and quantum strings,
Phys. Lett. B \textbf{265} (1991) 287-294.

\bibitem{Tseytlin:1991wr}
A.~A.~Tseytlin,
Duality and dilaton,
Mod. Phys. Lett. A \textbf{6} (1991) 1721-1732.

\bibitem{HZ}O.~Hohm and B.~Zwiebach,
Duality invariant cosmology to all orders in $\alpha'$,
Phys. Rev. D \textbf{100} (2019) no.12, 126011
[arXiv:1905.06963 [hep-th]].

\bb{21a}O.~Hohm and B.~Zwiebach,
Non-perturbative de Sitter vacua via $\alpha'$ corrections,
Int. J. Mod. Phys. D \textbf{28} (2019) no.14, 1943002
[arXiv:1905.06583 [hep-th]].


\bb{3}H.~Bernardo, R.~Brandenberger and G.~Franzmann,
O$(d,d)$ covariant string cosmology to all orders in $\alpha^{\prime}$,
JHEP \textbf{02} (2020) 178
[arXiv:1911.00088 [hep-th]].

\bb{4}J.~Quintin, H.~Bernardo and G.~Franzmann,
Cosmology at the top of the $\alpha'$ tower,
JHEP \textbf{07} (2021) 149
[arXiv:2105.01083 [hep-th]].

\bb{25}T.~Codina, O.~Hohm and B.~Zwiebach,
2D black holes, Bianchi I cosmologies, and $\alpha'$,
Phys. Rev. D \textbf{108} (2023) no.2, 026014
[arXiv:2304.06763 [hep-th]].

\bb{26}H.~Bernardo, R.~Brandenberger and G.~Franzmann,
String cosmology backgrounds from classical string geometry,
Phys. Rev. D \textbf{103} (2021) no.4, 043540
[arXiv:2005.08324 [hep-th]].

\bb{rost}C.~A.~N\'u\~nez and F.~E.~Rost,
New non-perturbative de Sitter vacua in $\alpha'$-complete cosmology,
JHEP \textbf{03} (2021), 007
[arXiv:2011.10091 [hep-th]].

\bibitem{Bieniek:2022mrv}
P.~Bieniek, J.~Chojnacki, J.~H.~Kwapisz and K.~A.~Meissner,
Stability of the nonperturbative O(D,D) de Sitter spacetime: The isotropic case,
Phys. Rev. D \textbf{106} (2022) no.10, 106009
doi:10.1103/PhysRevD.106.106009
[arXiv:2208.06010 [hep-th]].

\bb{27}P.~Bieniek, J.~Chojnacki, J.~H.~Kwapisz and K.~A.~Meissner,
 Stability of the de-Sitter spacetime. The anisotropic case, 
[arXiv:2301.06616 [hep-th]].

\bibitem{Song:2023txa}
L.~Song and D.~Chen,
Two non-perturbative \ensuremath{\alpha'} or loop corrected string cosmological solutions,
Chin. Phys. C \textbf{47} (2023) no.9, 095102
[arXiv:2306.07031 [hep-th]].

\bibitem{Bernardo:2021xtr}
H.~Bernardo, P.~R.~Chouha and G.~Franzmann,
Kalb-Ramond backgrounds in $\alpha'$-complete cosmology,
JHEP \textbf{09} (2021), 109
[arXiv:2104.15131 [hep-th]].

\bibitem{Gasperini:2023tus}
M.~Gasperini and G.~Veneziano,
Non-singular pre-big bang scenarios from all-order $\alpha'$ corrections,
JHEP \textbf{07} (2023) 144
[arXiv:2305.00222 [hep-th]].


\bibitem{Das:1986da}
S.~R.~Das and B.~Sathiapalan,
New Infinities in Two-dimensional Nonlinear $\sigma$ Models and Consistent String Propagation,
Phys. Rev. Lett. \textbf{57} (1986), 1511.

\bibitem{Codina:2023nwz}
T.~Codina, O.~Hohm and B.~Zwiebach,
On black hole singularity resolution in $D=2$ via duality-invariant $\alpha'$ corrections,
[arXiv:2308.09743 [hep-th]].
\bb{7}M. Gasperini and G. Veneziano, 
Pre - big bang in string cosmology, 
Astropart. Phys. {\bf 1} (1993) 317 
 [hep-th/9211021].

\bb{Wang1} P.~Wang, H.~Wu, H.~Yang and S.~Ying,
Non-singular string cosmology via $\alpha^{\prime}$ corrections,
JHEP \textbf{10} (2019) 263
[arXiv:1909.00830 [hep-th]].

\bb{Wang2} P.~Wang, H.~Wu, H.~Yang and S.~Ying,
Construct $\alpha^{\prime}$ corrected or loop corrected solutions without curvature singularities,
JHEP \textbf{01} (2020) 164
[arXiv:1910.05808 [hep-th]].

\bb{8}M. Gasperini and G. Veneziano, 
The Pre - big bang scenario in string cosmology, 
Phys. Rept. {\bf 373} (2003) 1 [hep-th/0207130].

\bb{9}M. Gasperini, {\it ``Elements of String Cosmology"},  Cambridge University Press (2007), ISBN 978-0-511-33229-6,
978-0-521-18798-5, 978-0-521-86875-4.

\bibitem{Buonanno:1998bi}
A.~Buonanno, T.~Damour and G.~Veneziano,
Pre - big bang bubbles from the gravitational instability of generic string vacua,
Nucl. Phys. B \textbf{543} (1999), 275-320
[arXiv:hep-th/9806230 [hep-th]].

\bibitem{Feinstein:2000ja}
A.~Feinstein, K.~E.~Kunze and M.~A.~Vazquez-Mozo,
Initial conditions and the structure of the singularity in pre - big bang cosmology,
Class. Quant. Grav. \textbf{17} (2000), 3599-3616
[arXiv:hep-th/0002070 [hep-th]].

\bb{gasp} M. Gasperini, 
On the initial regime of pre-big bang cosmology,
JCAP \textbf{09} (2017), 001
[arXiv:1707.05763 [gr-qc]].


\bibitem{Brustein:1994kw}
R.~Brustein and G.~Veneziano,
The Graceful exit problem in string cosmology,
Phys. Lett. B \textbf{329} (1994) 429-434
[arXiv:hep-th/9403060 [hep-th]].



\bibitem{Kaloper:1995ey}
N.~Kaloper, R.~Madden and K.~A.~Olive,
Axions and the graceful exit problem in string cosmology,
Phys. Lett. B \textbf{371} (1996) 34-40
[arXiv:hep-th/9510117 [hep-th]].



\bibitem{Easther:1995ba}
R.~Easther, K.~i.~Maeda and D.~Wands,
Tree level string cosmology,
Phys. Rev. D \textbf{53} (1996) 4247-4256
[arXiv:hep-th/9509074 [hep-th]].


\bibitem{Gasperini:1996in}
M.~Gasperini and G.~Veneziano,
Singularity and exit problems in two-dimensional string cosmology,
Phys. Lett. B \textbf{387} (1996), 715-720
[arXiv:hep-th/9607126 [hep-th]].

\bb{LL} L.D.~Landau and E.M.~ Lifshitz, {\it ``Mechanics"}, Third Edition, Pergamon Press (1976), p. 133.




\bb{Piazza}M. Gasperini, F Piazza and G. Veneziano, Quintessence as a runaway dilaton,
Phys. Rev. D \textbf{65} (2002) 023508 
[arXiv:gr-qc/0108016 [gr-qc]].


\bibitem{Damour:2002mi}
T.~Damour, F.~Piazza and G.~Veneziano,
Runaway dilaton and equivalence principle violations,
Phys. Rev. Lett. \textbf{89} (2002), 081601
[arXiv:gr-qc/0204094 [gr-qc]].


\bibitem{Damour:2002nv}
T.~Damour, F.~Piazza and G.~Veneziano,
Violations of the equivalence principle in a dilaton runaway scenario,
Phys. Rev. D \textbf{66} (2002), 046007
[arXiv:hep-th/0205111 [hep-th]].

\bibitem{Veneziano:2001ah}
G.~Veneziano,
Large N bounds on, and compositeness limit of, gauge and gravitational interactions,
JHEP \textbf{06} (2002), 051
[arXiv:hep-th/0110129 [hep-th]].

\bb{Bozza1}V. Bozza, M. Gasperini, M. Giovannini and G. Veneziano, Assisting pre-big bang
phenomenology through short lived axions, Phys. Lett. B {\bf 543} (2002) 14 [hep-ph/0206131].

\bb{Bozza2}V. Bozza, M. Gasperini, M. Giovannini and G. Veneziano, Constraints on pre big bang
parameter space from CMBR anisotropies, Phys. Rev. D {\bf 67} (2003) 063514 [hep-ph/0212112].

\bibitem{Gasperini:2003pb}
M.~Gasperini, M.~Giovannini and G.~Veneziano,
Perturbations in a nonsingular bouncing universe,
Phys. Lett. B \textbf{569} (2003), 113-122,
doi:10.1016/j.physletb.2003.07.028
[arXiv:hep-th/0306113 [hep-th]].


\bibitem{Kunze:1999xp}
K.~E.~Kunze and R.~Durrer,
Anisotropic ``hairs" in string cosmology,
Class. Quant. Grav. \textbf{17} (2000) 2597-2604
[arXiv:gr-qc/9912081 [gr-qc]].

\bb{Quintin}C.~Ganguly and J.~Quintin,
Microphysical manifestations of viscosity and consequences for anisotropies in the very early universe,
Phys. Rev. D \textbf{105} (2022) no.2, 023532
[arXiv:2109.11701 [gr-qc]].

\bb{24}M. Gasperini, Looking back in time beyond the big bang, 
Mod. Phys. Lett. A {\bf 14} (1999) 1059
[gr-qc/9905062].


\bibitem{Brandenberger:2012aj}
R.~H.~Brandenberger and J.~Martin,
Trans-Planckian Issues for Inflationary Cosmology,
Class. Quant. Grav. \textbf{30} (2013), 113001
doi:10.1088/0264-9381/30/11/113001
[arXiv:1211.6753 [astro-ph.CO]].


\end{thebibliography}
\end{document}